\documentclass[epj]{svjourblank}

\usepackage{graphics}
\usepackage{amsmath}
\usepackage{amsfonts}
\usepackage{amssymb}
\usepackage{fancyhdr}

\renewcommand{\vec}[1]{\mbox{\boldmath$#1$}}

\newcommand{\w}{O}

\newcommand{\uc}{\xi}

\begin{document}
\title{A General Theory of Non-equilibrium Dynamics of Lipid-protein Fluid Membranes}

\author{Michael A. Lomholt\thanks{\emph{e-mail:} mlomholt@fysik.sdu.dk}
\and Per L. Hansen\thanks{\emph{e-mail:} lyngs@memphys.sdu.dk}
\and Ling Miao\thanks{\emph{e-mail:} miao@memphys.sdu.dk; author of correspondence}}

\institute{The MEMPHYS Center for Biomembrane Physics, Physics Department, University of
Southern Denmark,\\ DK-5230 Odense M, Denmark}

\date{}

\abstract{
We present a general and systematic theory of non-equilibrium dynamics of multi-component
fluid membranes, in general, and membranes containing transmembrane proteins, in particular.  Developed
based on a minimal number of principles of statistical physics and designed to be a meso/macroscopic-scale
effective description, the theory is formulated in terms of a set of equations of hydrodynamics and linear
constitutive relations.  As a particular emphasis of the theory, the equations and the constitutive
relations address both the thermodynamic and the hydrodynamic consequences of the unconventional material
characteristics of lipid-protein membranes and contain proposals as well as predictions which have not
yet been made in already existed work on membrane hydrodynamics and which may have experimental relevance.
The framework structure of the theory makes possible its applications to a range of non-equilibrium
phenomena in a range of membrane systems, as discussions in the paper of a few limit cases demonstrate.    
}
\maketitle

\section{Introduction}
\label{sec:intro}
Lipid-protein fluid membranes are the most essential structural element in biological
cells, defining boundaries of the cells and intracellular organelles.  They are also
one of the important functional elements, actively participating in many cellular
processes.  Each biological membrane is composed of a core bilayer of amphiphilic
lipids, in which transmembrane proteins are embedded and with which peripheral proteins are
associated.  In its functional state, the membrane is fluid, allowing the constant movement
and organization of the constituent mole\-cules within the structure, and its two-dimensional
geometry deforms easily in connection with its function, requiring energies comparable to
thermal energy only.  Moreover, the membrane constantly exchanges material and energy with
its environment; active transport of small solute molecules across the membrane takes place
constantly, carried out by membrane proteins that require external energy sources, such as
chemical energy provided by ATP, electrochemical energy stored in cross-membrane proton 
gradients, or light.  From the point of view of statistical physics, it is obvious that such
a functioning membrane should be treated as a non-equili\-brium system and that the dynamics
of the membrane is intimately coupled to the dynamics of the bulk fluids within which it is
embedded.
  
During the last three decades, investigations, characterizations and understanding of the
physical properties and behaviour of such membranes and simpler model membranes have fueled
both the development of statistical physics of soft condensed matter in general and the development 
of membrane biophysics in particular, because these systems have posed many issues challenging
the traditional framework of statistical physics and also because it has become more and more
appreciated that understanding of the physics of the membranes can shed light on their biological
functions \cite{greenbooks}.  

For obvious reasons, most of the development of membrane statistical physics has focussed only
on the equilibrium, static aspect of the membrane systems.  To the purpose of a more complete physical
description of biological membranes, however, their non-equilibrium behaviour must be investigated
and described.  The work presented in this paper is our attempt at taking a step in that direction.   

Some recent experimental as well as theoretical studies of simpler model systems of biological
membranes have particularly motivated our work \cite{french}.  In the model systems, a single type
of transmembrane protein, which can be externally driven to actively transport small ions across
the membrane, was reconstituted into a core lipid bilayer at various concentrations.  By turning
on or off the external driving force, the lipid-protein composite membrane was then set in either
a non-equilibrium or an equilibrium state, and the non-equilibrium dynamics of the membrane
conformation was then investigated experimentally.  Theories, in the form of equations of dynamics,
were also developed to describe or interpret the experimental observations.  The formulation of the
theories appears, however, entirely intuition based, which makes its validity rather opaque and its
generalization to other lipid-protein composite membranes difficult.  We have, therefore, made it
our goal to develop a more general and systematic theoretical description of lipid-protein composite
membranes in non-equilibrium states, believing that such a theoretical framework will become useful
as more and more experimental studies of such systems will be done with their results to be interpreted
theoretically.  

In fact, a conceptual basis for developing a general description already exists, and it consists
of a minimal number of principles of statistical physics: the basic laws of thermodynamics, formulated
as a balance equation of entropy, the assumption of local thermodynamic equilibrium in a non-equilibrium
system and the dynamic formulation of laws of conservation of mass, momentum and energy, and if 
relevant, angular momentum.\footnote{These conservation laws are sufficient if we limit ourselves to
the non-equilibrium dynamics which does not involve non-conserved physical variables.}  However,
the explicit expressions and implementation of these principles in the case of lipid-protein
composite membranes, which in the end result in a set of clearly formulated equations of
(hydro)dynamics, must be developed systematically and unambiguously to reflect the physical
characteristics pertaining to the membranes.

At the level of continuum descriptions, a membrane is modelled effectively as an interface separating
two bulk fluids.  Its physical characteristics, however, distinguish it from a simple interface
separating two coexisting fluids.  Firstly, its molecular composition and material structure differ
significantly from those of the two bulk fluids with which it is in contact, and the motion and
organization of its constituent molecules within its interface structure provide a set of interesting
phenomena.  Thus, the material content, and associated with it, the energy content as well as momentum
content of the membrane should in general be taken into account.  Secondly, its energy (or, thermodynamics)
depends not only on its surface area, but also on its {\em local} surface geometry, or curvatures
\cite{helfrich,evans}.  Thirdly, its properties with regard to processes of transport of material
and energy are in general markedly different from those of the bulk fluids.  A good example is the
much slower diffusion across the membrane than in the bulk fluids of polar molecules.  Thus, the
dissipations associated with such transport processes should not be neglected {\em a priori}, as is
canonically done when non-equilibrium dynamics involving a conventional interface is considered.
In fact, there has been experimental evidence suggesting that a membrane should not be considered
as dissipationless in its non-equilibrium state \cite{yeung95,seifert93,miao02}.  Finally, the typical
constituent molecules of a lipid-protein composite membrane -- the smaller amphiphilic lipid molecules
and much larger transmembrane protein molecules -- differ significantly in their molecular chemistry
and structures.  Consequently, their interactions with the contacting bulk fluids may
also be expected to differ.  Our work, in essence, consists in recognizing these unconventional
characteristics, clarifying the basic conceptual issues that inevitably rise from considering
them, and finally developing an unambiguous description of these characteristics in the
context of non-equilibrium dynamics of the system.

To be sure, studies concerning fundamental descriptions of non-equilibrium dynamics of fluid surfaces
and interfaces began already three decades ago \cite{defay,bedeaux76}.  For example Bedeaux {\em et al.}
developed for conventional interfaces a formulation of non-equilibri\-um dynamics by taking into account
the energy and momentum content of the interface, from which a connection to equilibrium interface
thermodynamics can be made \cite{bedeaux76}.  More recently, a theory for the non-equilibrium dynamics
of a two-component surfactant interface separating air and water was presented in the form of a set
of dynamic equations, with an emphasis on nonlinear phenomena where phase separation and surface
deformations are coupled \cite{onuki93}.  In this theory, the mass content and the composition of the
interface were explicitly taken into account, and the dependence of the interface thermodynamics on
the curvatures was also introduced.  The derivation of the dynamic equations was, however, more
intuitive than systematic, and certain assumptions were made, which were neither necessary nor
justified.  Moreover, the transverse transport processes were not discussed.  A generalization
of this theory to lipid-protein composite membranes did not appear straightforward.

It should be helpful to the reader that we at the outset describe more specifically the system
and its non-equilibri\-um state that we have in mind when developing our theory.  The essential
picture is sketched out in Fig.~1(a).  The system we consider consists of a multi-component
membrane of lipid-protein composite, which has a nanometer thickness and which assumes a quasi
two-dimensional geometrical shape, and of two aqueous fluids, which the membrane is in contact with.
The two fluids can be identical in their chemical compositions and equilibrium thermodynamics,
or distinctly different, as in the case of two fluids under conditions of phase separation.
In order to keep the theory general, we allow the membrane geometry to be arbitrary, and we do not
specify quantitatively the number of molecular species constituting the membrane, nor their exact
chemical structures, although the cases where the protein components are transmembrane proteins are
of particular interest to us.  We also assume that the aqueous fluids may contain more than one types
of solute molecules.

\begin{figure}
\vspace*{.5cm}
\centerline{
\resizebox{8cm}{!}{
  \includegraphics{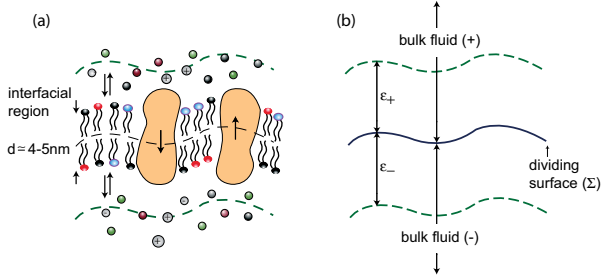}  
}}
\caption{A schematic sketch of the membrane-fluid system under our consideration.  Fig.~1(a) depicts 
the interfacial region in the real system, showing a membrane composed of bilayer-forming lipids
and transmembrane proteins in contact with two fluids containing small solutes.  Fig.~1(b) illustrates
the representation of the interfacial region in the corresponding Gibbs model system.}
\end{figure}

We then consider a general non-equilibrium state, where chemical, thermal and mechanical
gradients exist both in the directions tangent to the membrane surface and in the direction
transverse to the membrane, and where chemical reactions may take place on the membrane.  But
in the scope of this paper, we will limit ourselves to situations where the length scales over
which the assumption of local thermodynamic equilibrium is valid are larger, though not by
orders of magnitude, than the thickness of the membrane.  The corresponding time scales may
be expected to approach the mesoscopic or above.  Given the typical time resolutions of around
milliseconds of most experimental techniques used to study membranes, our theory will have
a reasonably large range of applicability in terms of length and time scales.   It follows then
that the membrane is assumed to be in thermodynamic equilibrium with the fluids which it is in
immediate contact with.\footnote{Formulation of thermodynamics in situations where this assumption
is not valid was first discussed in Ref.~\cite{defay} in the context of ``dynamic surface tension." }  

We would also like to briefly state the philosophy which we have followed when developing the
theory.  Although the membrane-fluid systems we consider are on microscopic length scales highly
inhomogeneous in molecular distributions and transport properties, our theory is not concerned 
with accurately describing the details of the microscopic-scale inhomogeneity, but aims at
describing the inhomogeneity only through those effects that may be observed at
mesoscopic or macroscopic scales.   Following this philosophy, we have employed an idea originated
from Gibbs \cite{gibbs}: the idea of replacing a real, inhomogeneous membrane-fluid system with a model
system of two homogeneous bulk fluids separated by an infinitely thin dividing surface and relating
to the dividing surface all the excess thermodynamic and hydrodynamic contributions that are required
by an equivalence between the thermodynamic and hydrodynamic behaviour of the two systems.  This idea
is also illustrated graphically by Fig.~1.  For later convenience, we will dub this model system
``the Gibbs system."  

The paper is organized as follows.  In Section~\ref{sec:geom} the necessary surface differential
geometry is briefly reviewed.  In Section~\ref{sec:therm} the concept of the surface thermodynamics
of the membrane is unambiguously defined based on the idea of the Gibbs model system, and a general
expression of the surface thermodynamics is derived and discussed.  In addition, a general identity
relating different thermodynamic variables associated with the membrane surface is derived, which
takes into account the salient mechanical characteristics of membranes.  To our knowledge, this is
the first derivation of such an identity.  An assumption made in this Section is that intrinsic
orientational degrees of freedom are not relevant.   In Section~\ref{sec:conslaws} a description of
the hydrodynamics of the whole membrane-fluid system is formulated by defining relevant ``bulk"
and ``surface" hydrodynamic variables and by establishing equations of dynamics for all of the
hydrodynamic variables based on the fundamental conservation laws.  The hydrodynamic description
is limited to cases, where systems under consideration have no intrinsic angular momenta.  In
Section~\ref{sec:eprod}, the results from Section~\ref{sec:therm} and Section~\ref{sec:conslaws},
combined with the assumption of local thermodynamic equilibrium  are used to derive the entropy
production, based on which conjugate pairs of general thermodynamic/hydrodynamic force and dissipative
current are identified.  Constitutive relations for the linear non-equilibrium dynamics of the system
are derived in Section~\ref{sec:consrel}.  In particular, some of these constitutive relations, which
under appropriate conditions reduce to those that appear familiar, make predictions about new mechanisms
governing various dissipative processes in the membrane.  The complete set of constitutive relations,
together with the equations of dynamics, thus provide a closed formulation of the hydrodynamics of
the whole membrane-fluid system.  In Section~\ref{sec:limitcases}, a number of limit cases of the
general theory, which have practical relevance, are discussed, in order that the theory, having
been presented in a general and formulistic way, also be seen from a practical point of view.
Finally, in the concluding section, Section~\ref{sec:discussion}, the theory is discussed within
the context of its applications and its connections to experimental measurements.  In order that
a technical point be made clear, which we expect will often be encountered in applications of the
theory, a short appendix is also attached.  

\section{Differential geometry of surfaces}
\label{sec:geom}
In the continuum theory which we will develop in this paper, a membrane is structurally described
as a two-dimensional surface.  In this section we briefly review, main\-ly to establish the notation,
the mathematical language of two-dimensional differential geometry, which will be used to describe
membrane geometry.  A more comprehensive introduction can be found in, for instance, Refs.~\cite{spivak,aris}.

The dynamic shape of the surface is represented by a space-vector function $\vec{R}=\vec{R}(\uc^1,\uc^2,t)$.
The variables $\uc^1$ and $\uc^2$ are internal coordinates corresponding to a parame\-trization of the
surface and $t$ represents time.  At each point on the membrane a basis for three-dimensional vectors
can be established.  Two of them are the tangential vectors
\begin{equation}
\vec{t}_\alpha \equiv \partial_\alpha\vec{R} \equiv \frac{\partial \vec{R}}{\partial \uc^\alpha}\ ,
\end{equation}
where $\alpha=1,2$, and the third is a unit vector normal to the surface,
\begin{equation}
\label{eq:normaldef}
\vec{n} \equiv \frac{\partial_1\vec{R}\times\partial_2\vec{R}}{|\partial_1\vec{R}\times\partial_2\vec{R}|}\ .
\end{equation}

The local geometry of the surface is characterized by two surface tensors, the metric tensor and the
curvature tensor.  The local metric tensor is defined by
\begin{equation}
g_{\alpha\beta} \equiv \partial_\alpha\vec{R}\cdot\partial_\beta\vec{R}\ .
\end{equation}
It has an inverse, $g^{\alpha\beta}$, which satisfies by definition
\begin{equation}
g^{\alpha\beta}g_{\beta\gamma} = \delta^\alpha_\gamma\;,
\end{equation}
where $\delta^\alpha_\gamma$ is the Kronecker delta and where the repeated Greek superscript-subscript
indices imply summation following the Einstein summation convention.  The metric tensor and its inverse
are used to raise and lower Greek indices as in the following example:
\begin{equation}
\vec{t}^\alpha=g^{\alpha\beta}\vec{t}_\beta\;,\quad \vec{t}_\alpha=g_{\alpha\beta}\vec{t}^\beta\;.
\end{equation}
The curvature tensor $K_{\alpha\beta}$ is defined via the second derivatives of the surface shape
function,
\begin{equation}
K_{\alpha\beta} = \vec{n} \cdot \partial_\alpha\partial_\beta\vec{R}\;.
\end{equation}
From $K_{\alpha\beta}$ the scalar mean curvature $H$ and Gaussian curvature $K$ can be obtained:
\begin{align}
H&=\frac{1}{2}g^{\alpha\beta}K_{\alpha\beta}\ ,\\
K&=\det g^{\alpha\beta}K_{\beta\gamma}\ .
\end{align}

Two other tensors will also be introduced here for later convenience,
\begin{equation}
\varepsilon_{\alpha \beta} \equiv \epsilon_{\alpha \beta} \sqrt{g} \;, \quad \quad \quad
\varepsilon^{\alpha \beta} \equiv \epsilon^{\alpha \beta}/ \sqrt{g} \;,
\end{equation}
where $\epsilon_{\alpha \beta} = \epsilon^{\alpha \beta}$ with $\epsilon_{11}= \epsilon_{22}= 0$
and $\epsilon_{12} = - \epsilon_{21} = 1$ are relative tensors, and $g=\det g_{\alpha\beta}$ is
the determinant of the metric tensor.

Expressions of covariant/contravariant differentiations of vector and tensor functions defined
on the surfaces are facilitated by the use of the Christoffel symbols, $\Gamma^\gamma_{\alpha\beta}$.
One instance, which will become particularly useful later, is the covariant differentiation of
a surface vector function, $\vec{w}=w^\alpha\vec{t}_\alpha$, given by
\begin{equation}
D_\alpha w^\beta = \partial_\alpha w^\beta + w^\gamma \Gamma^\beta_{\gamma\alpha}\;.
\end{equation}
The Christoffel symbols can also be defined as certain combinations of the derivatives of the
metric tensor, \linebreak 
namely,
\begin{equation}
\Gamma^\gamma_{\alpha\beta} = \frac{1}{2} g^{\gamma\delta} \left(
                           \frac{\partial g_{\delta \alpha}}{\partial \xi^{\beta}}
                           + \frac{\partial g_{\beta \delta}}{\partial \xi^{\alpha}}
                           -  \frac{\partial g_{\alpha \beta}}{\partial \xi^{\delta}} \right) \;.
\end{equation}
It follows that the covariant divergence of $w^\alpha$ can be written as 
\begin{equation}
D_\alpha w^\alpha=\frac{1}{\sqrt{g}}\partial_\alpha \left(\sqrt{g}w^\alpha\right)\;.
\end{equation}

Finally, the area of a local differential element of the surface is given by
\begin{equation}
dA=\sqrt{g}d\uc^1d\uc^2\;,
\end{equation}
an expression which will be repeatedly used in surface integrals.  

\section{Membrane thermodynamics}
\label{sec:therm}
An important component of a general description of non-equilibrium dynamics of the membrane-fluid system
sketch\-ed out in Introduction is an appropriate formulation of the equilibrium thermodynamics of the
system.  In this section we will present such a formulation.  Two points which pertain to the
system and which have been dealt with in the formulation, may already be mentioned here.  Firstly,
to describe the thermodynamic effects arising from the microscopic inhomogeneity inherent in the
system, we have employed the idea of the Gibbs model system.   A subtle issue of principle arises,
however, when Gibbs' idea is applied to the membrane system.  In the case of a conventional capillary
interface, whose mechanical property is entirely described
macroscopically by a surface tension experimentally measurable, the position of the Gibbs dividing
surface is {\em uniquely} determined by the thermodynamic -- including mechanical -- equivalence between
the real and the Gibbs systems \cite{defayprigogine,kondo}.  In the case of the membrane system, whose
macroscopic mechanical properties include not only a ``membrane tension," but also ``membrane bending
moments," the thermodynamic equivalence between the real and the Gibbs systems still
leaves the position of the Gibbs dividing surface free to be chosen in principle.  With a view of
making connection to experimental, in particular mechanical, characterizations of membrane systems,
where the geometrical profile of a membrane surface is resolved with optical resolutions 
\cite{videomicroscopy}, we do not define quantitatively the position of the theoretical surface, but
will work under the assumption that a position can be chosen to be consistent with the experimentally
determined one.

Secondly, related to the significance of membrane bending moments in the effective description
of a general membrane system \cite{helfrich,evans}, the thermodynamic free energies of the membrane
systems under our consideration must depend on {\em local} geometrical properties such as the principal
curvatures.  We recognize that, due to such dependence, the free energies no longer scale homogeneously
with the size of the membrane and that, consequently, the Gibbs adsorption equation in its
canonical form \cite{defayprigogine} no longer exists.  However, a certain statement, in the form of
an identity, can still be made about how the different surface thermodynamic variables are related.
The derivation of this identity will also be given in this section.  

\subsection{The basic equation}
\label{subsec:basicequation}
In order to develop a local description of the equilibrium thermodynamics of the membrane system,
it is necessary that we consider a small cell, or a volume element, ${\bar{\Sigma}}$, of the whole
membrane-fluid system.  A sketch of the cell is given in Fig.~2.  The spatial extensions of the cell
are defined by a base area element $\Sigma$ on the dividing surface, which spans in a chosen internal
coordinate space a fixed rectangle with corners at $(\uc^1-\Delta\uc^1/2,\uc^2\pm\Delta\uc^2/2)$ and
$(\uc^1+\Delta\uc^1/2,\uc^2\pm\Delta\uc^2/2)$, and by a height $\epsilon^+$ above and a height $\epsilon^-$
below the dividing surface in the direction of the surface normal vector $\vec{n}$.   The local geometry
of the surface is described by $\vec{R}(\uc^1,\uc^2)$.  $\epsilon^+$ and $\epsilon^-$ are chosen such
that the physical characteristics of the system at $\epsilon^+$ and $\epsilon^-$ reach those of the
two homogeneous bulk fluids.  The cell is assumed to be in thermal equilibrium with a uniform temperature,
and it is also assumed to  be in mechanical equilibrium, although this does not necessarily imply a
uniform pressure across the whole cell due to the fact that the membrane interface may be curved.
Moreover, in order to include mass motion in the formulation of thermodynamics, the cell is considered
to be in uniform motion characterized by a velocity $\vec{v}$. 

\begin{figure}
\vspace*{.5cm}
\centerline{
\resizebox{8cm}{!}{
  \includegraphics{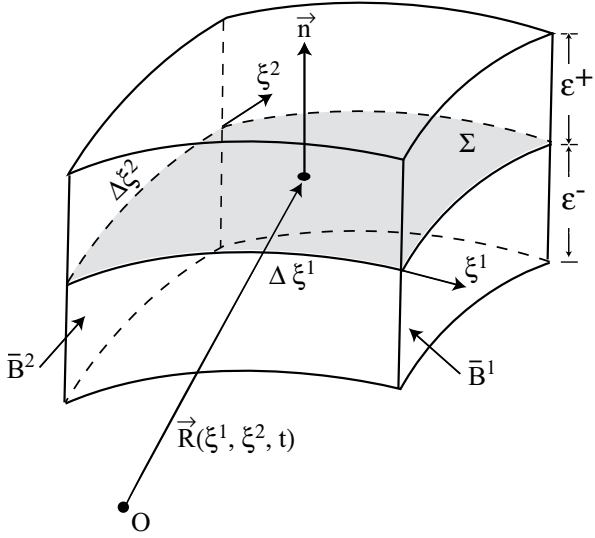}  
}}
\caption{An illustration of the cell, or volume element, ${\bar{\Sigma}}$.  $\bar{B}^{1}$ and $\bar{B}^{2}$
refer to the side surfaces of the cell. }
\end{figure}

Regarding the state of chemical equilibrium in the cell, particular considerations are needed.  It is
well known that a typical membrane appears impermeable to polar mole\-cules such as small ions on the time
scales of seconds \cite{no_permeation} and that the reorientations of the constituent amphiphilic lipid
and protein molecules from one side of the membrane surface to the other are not observed on similar
time scales \cite{flipflop_1}.  To take into account this fact, therefore, we introduce an operational
definition of {\em species}, which is broader than the canonical definition based on the chemical nature
of a molecular component: solution molecules of the same chemical structure, but found to be located on
the two different sides of the dividing surface will be counted as two different species if their transport
across the membrane is a slow process; lipid, or protein molecules having the same chemistry, but oriented
in the two opposite directions across the membrane will also be considered as two different species.
The equilibrium state which we will develop a thermodynamic description for corresponds to a state
where each species is in its chemical equilibrium, but where there is no chemical equilibrium, in
the {\em quasi-static} sense, between any two species of the same molecular chemistry.  To maintain
the generality of the theory, we will also allow for in the system the presence of molecular components
which do reach chemical equilibrium across the membrane in the same quasi-static sense.  Each such
component constitutes a single species by our definition. 

We denote the energy content in this cell by $E_{\bar{\Sigma}}$.  Following the formal structure
of thermodynamics, we assume that $E_{\bar{\Sigma}}$ depends on only a few state variables: the
molecular numbers of the different species, $N_{A,{\bar{\Sigma}}}$, where ``$A$" is an index
labelling the species; the entropy $S_{\bar{\Sigma}}$; the momentum $\vec{P}_{\bar{\Sigma}}$;
and finally, the variables characterizing the shape and size of the cell, namely, the heights
$\epsilon^\pm$, which are assumed to be fixed, and derivatives of the shape field
$\vec{R}(\uc^1,\uc^2)$ (such as $H$ and $K$).\footnote{In making this assumption, we exclude
from our considerations cases where intrinsic orientational degrees of freedom and their
independent rotational motion are relevant.  An extension which does include those effects
is given in Ref.~\cite{michael_thesis,stress}.}  The reason that only derivatives of $\vec{R}$,
not $\vec{R}$ itself, are allowed is that the thermodynamics should be invariant under rigid
translations, provided that the translational-symmetry-breaking effects such as the gravitational
force are taken into account as external forces.  Based on this assumption, we can therefore express,
for the chosen cell, the first and second laws of thermodynamics as a complete differential of
$E_{\bar{\Sigma}}$ with respect to all the relevant state variables: 
\begin{align}
\delta E_{\bar{\Sigma}}= \;&\vec{v}\cdot\delta\vec{P}_{\bar{\Sigma}}+\sum_A\mu^A\delta N_{A,{\bar{\Sigma}}}
                      +T\delta S_{\bar{\Sigma}}\nonumber\\
&-\vec{F}_{\bar{\Sigma}}\cdot\delta\vec{R}
                      +\partial_\alpha\left(\vec{S}^\alpha_{\bar{\Sigma}}\cdot\delta\vec{R}\right)\;,\label{cell}
\end{align}
where we have already identified the partial derivatives with respect to $\vec{P}_{\bar{\Sigma}}$,
$N_{A,{\bar{\Sigma}}}$, and $S_{\bar{\Sigma}}$ with the velocity $\vec{v}$, the chemical potential
for species $A$, and the temperature $T$ of the cell. 

In the last two terms, $\vec{F}_{\bar{\Sigma}}$ is a regular quantity and carries the significance
of a physical force, whereas $\vec{S}^\alpha_{\bar{\Sigma}}$ contains differential operators and
is related to surface stresses.  From a mathematical point view, their presence is not
difficult to understand, as they represent the variation in the energy function resulted from
any variation in the shape field, $\vec{R}(\uc^1,\uc^2)$, and in turn, variations in its derivatives.
Seen from a more physical point of view, the two terms must describe the mechanical work done on
the cell when the shape of the dividing surface is changed.  To illustrate how their functional
forms arise from their mechanical origins is not a trivial issue, and is discussed in a separate
paper \cite{stress}.  However, mechanical interpretations of $\vec{F}_{\bar{\Sigma}}$ and
$\vec{S}^\alpha_{\bar{\Sigma}}$ will be made a bit later in the paper, clarifying the reason for
expressing the work in those particular functional forms. 

To reformulate Eq.~(\ref{cell}) by the use of the Gibbs model system, we introduce bulk volume densities
for extensive quantities of the cell: ${\bar e}^\pm$, $\vec{\bar p}^\pm$, ${\bar s}^\pm$ and
${\bar n}^\pm_A$.  These can be expressed as functions of intensive thermodynamic variables $\vec{v}$,
$T$ and $\mu^A$'s and are assumed to represent the known thermodynamic behaviour of the homogeneous
bulk fluids in the following sense: 
\begin{equation}\label{eq:firstbulk}
\delta{\bar e}^\pm=\vec{v}\cdot\delta\vec{\bar p}^\pm+T\delta{\bar s}^\pm+\sum_A\mu^A\delta{\bar n}_{A}^\pm\;,
\end{equation}
and
\begin{equation}\label{eq:GDbulk}
{\bar e}^\pm=\vec{v}\cdot\vec{\bar p}^\pm+T{\bar s}^\pm+\sum_A\mu^A{\bar n}_{A}^\pm-{p}^\pm\;,
\end{equation}
where $p^\pm$ denotes the bulk function of velocity, temperature and chemical potentials that
corresponds to pressure.  

We can now define the excess area densities of extensive quantities, which must be associated
with the dividing surface according to the construction of the Gibbs model system: 
\begin{align}
e & =\frac{1}{A_\Sigma}\left(E_{\bar{\Sigma}}-V^+_{\bar{\Sigma}}{\bar e}^+
    -V^-_{\bar{\Sigma}}{\bar e}^-\right)\;,\\
\vec{p} & = \frac{1}{A_\Sigma}\left(\vec{P}_{\bar{\Sigma}}-V^+_{\tilde\Sigma}\vec{\bar p}^+
          -V^-_{\bar{\Sigma}}\vec{\bar p}^-\right)\;,\\
s & = \frac{1}{A_\Sigma}\left(S_{\bar{\Sigma}}-V^+_{\bar{\Sigma}}{\bar s}^+
      -V^-_{\bar{\Sigma}}{\bar s}^-\right)\;,\\
n_A & = \frac{1}{A_\Sigma}\left(N_{A,{\bar{\Sigma}}}-V^+_{\bar{\Sigma}}{\bar n}_A^+
        -V^-_{\bar{\Sigma}}{\bar n}_A^-\right)\;,
\end{align}
where $A_\Sigma$ is the area of the surface element $\Sigma$, and $V^\pm_{\bar{\Sigma}}$ represents
the volume of the cell part which is above/below the dividing surface, respectively.  It follows
that
\begin{equation}\label{eq:peqofstate}
\vec{p}=\rho\vec{v}\;,
\end{equation}
where $\rho=\sum_A m^{A}n_{A}$ is the excess mass density.

Using these excess quantities, together with 
\begin{align}
V^\pm_{\bar{\Sigma}}&=A_\Sigma\left(\epsilon^\pm\mp(\epsilon^\pm)^2H+\frac{1}{3}(\epsilon^\pm)^3K\right)\;,\\
A_\Sigma&=\Delta\uc^1\Delta\uc^2\sqrt{g}\;,
\end{align}
where we have assumed that the radii of curvature are bigger than $\epsilon^\pm$, we finally arrive at
the following reformulation of Eq.~(\ref{cell}),
\begin{align}
\delta\left(\sqrt{g} e\right)=&\;\vec{v}\cdot\delta\left(\sqrt{g}\vec{p}\right)+T\delta\left(\sqrt{g} s\right)+\sum_A\mu^A\delta\left(\sqrt{g} n_A\right)
                      \nonumber\\
&                 - \sqrt{g}\vec{f}_{\rm rs}\cdot\delta\vec{R} 
                  +\sqrt{g}D_\alpha\left(\vec{S}^\alpha\cdot\delta\vec{R}\right)\;.
\label{eq:varint}
\end{align}
Vector quantities $\vec{f}_{\rm rs}$ and $\vec{S}^\alpha$ are defined by
\begin{align}
\vec{f}_{\rm rs} & \equiv \frac{\vec{F}_{\bar{\Sigma}}}{A_\Sigma}-D_\alpha\vec{B}^\alpha_+-D_\alpha\vec{B}^\alpha_-\;,\\
\vec{S}^\alpha & \equiv \frac{1}{A_\Sigma}\vec{S}^\alpha_{\bar{\Sigma}}-\vec{B}^\alpha_+
              -\vec{B}^\alpha_- + \vec{n}\left( {C}^{\alpha\beta}_{+} +
              {C}^{\alpha\beta}_{-} \right)\partial_\beta \;,
\end{align}
where
\begin{align}
\vec{B}^\alpha_\pm \equiv & \; - \epsilon^\pm p^\pm \vec{t}^\alpha \pm p^\pm(\epsilon^\pm)^2
                  \left( Hg^{\alpha\beta}-\frac{1}{2}K^{\alpha\beta}\right) \vec{t}_\beta\nonumber\\
&- D_\beta C^{\alpha\beta}_{\pm} \,\vec{n} \;,\\
{C}^{\alpha\beta}_\pm \equiv & \;\mp(\epsilon^\pm)^2p^\pm\frac{1}{2}g^{\alpha\beta}
+\frac{1}{3}(\epsilon^\pm)^3p^\pm\left(2Hg^{\alpha\beta}-K^{\alpha\beta}\right)\;.
\end{align}
Note that the physics represented by Eq.~(\ref{eq:varint}) should be independent of any specific numerical
values of $\epsilon^{\pm}$.  

Based on Eq.~(\ref{eq:varint}) the mathematical area element $\Sigma$ on the dividing surface may be viewed
as an effective surface system which has its own thermodynamic properties and which interacts with its
``surroundings."  In this effective picture, the last two terms in Eq.~(\ref{eq:varint}) -- related to the
mechanical-work terms in Eq.~(\ref{cell}) -- may be interpreted as the work done on the effective surface
system by its surroundings: In Section 5, it will become clear that $\vec{f}_{\rm rs}$ must balance the
{\em effective} force exerted on $\Sigma$ by sources external to the dividing surface under mechanical
equilibrium.  Similarly, $\sqrt{g}D_\alpha\left(\vec{S}^\alpha\cdot\delta\vec{R}\right)$ will be shown
in Eq.~(\ref{eq:jer}) to represent the work done on $\Sigma$ by the rest of the dividing surface.  But, this
interpretation can already be made plausible here by noting that
$\int_{\Sigma}d^2\uc\;\sqrt{g}D_\alpha\left(\vec{S}^\alpha\cdot\delta\vec{R}\right)$ 
can actually be rewritten as an integral over the boundary $\partial\Sigma$ of the area element $\Sigma$
\begin{equation}
\int_{\Sigma}\;d^2\uc\;\sqrt{g}D_\alpha\left(\vec{S}^\alpha\cdot\delta\vec{R}\right)
=\int_{\partial\Sigma}d s\;\nu_\alpha\vec{S}^\alpha\cdot\delta\vec{R}\;,
\end{equation}
where $s$ is the arc length and $\nu_\alpha\vec{t}^\alpha$ is a unit normal vector pointing away from
the boundary $\partial\Sigma$.  

There is in fact a connection between $\vec{f}_{\rm rs}$ and $\vec{S}^\alpha$.  If $\vec{S}^\alpha_{(0)}$
is used to represent all the contributions in $\vec{S}^\alpha$ that do not involve differential operators,
then it can be seen from Eq.~(\ref{eq:varint}) that the following relationship must be satisfied, as a
consequence of the invariance of the thermodynamics under rigid translations: 
\begin{equation}
\label{eq:forcestressrel}
\vec{f}_{\rm rs} = D_\alpha\vec{S}^\alpha_{(0)} \;.
\end{equation}
$\vec{f}_{\rm rs}$ accounts for the total mechanical force exerted on the effective surface element
$\Sigma$ by the rest of the effective surface.  In what follows, $\vec{f}_{\rm rs}$ will be referred to
as the ``restoring force."  It is clear that $\vec{S}^\alpha_{(0)}$ should be identified as the surface
stress tensor as defined in \cite{kralchevsky94,capovilla02}.  

The above interpretations make clear the reason for organizing the geometry-dependent work explicitly
into the two particular functional terms in Eq.~(\ref{cell}).  We would like to note also that
these interpretations can be obtained in a more physically intuitive and direct way by formulating
the mechanical work explicitly, once a model of the mechanical behaviour of the inhomogeneous cell
is given.  $\vec{F}_{\bar{\Sigma}}$ and $\vec{S}^\alpha_{\bar{\Sigma}}$, or $\vec{f}_{\rm rs}$ and
$\vec{S}^\alpha$, can be identified with the mechanical model quantities.  But, we defer the discussion
of that topic to another paper \cite{stress}.  

It can be seen easily that Eq.~(\ref{eq:varint}) does not define $\vec{S}^\alpha$ uniquely.  An addition
to it of the following type, for instance,
\begin{equation}
\varepsilon^{\alpha\beta} \left( \partial_\beta\vec{V} + \vec{V}\partial_\beta \right) \;,
\end{equation}
where $\vec{V}$ is an arbitrary vector, does not change the mechanical work at all.  This seemingly
mathematical point implies in fact a non-trivial physical statement: the theoretical characterization
of the mechanical behaviour of a membrane system in terms of an effective surface stress tensor and 
surface bending-moment tensor is not unique, as opposed to the ``belief" implied in the canonical
description of membrane mechanics \cite{evansskalak}.  We will discuss and clarify this issue elsewhere
\cite{stress}.  

Eq.~(\ref{eq:varint}) thus provides the basic equation of the membrane surface thermodynamics and will
be used later in our description of non-equilibrium dynamics of the membrane system.  Once an explicit
functional form of $e$ and the values of the surface thermodynamic variables, $\vec{p}$, $s$, $n_A$'s,
and $\vec{R}$ are assumed to be known, Eq.~(\ref{eq:varint}), or its integrated form $E=\int_{\rm M} d A\;e$,
can be used to determine the other {\em physically measurable} thermodynamic variables of the surface
as follows:
\begin{align}
\label{derivatives}
\vec{v}&=\frac{1}{\sqrt{g}}\left.\frac{\delta E}{\delta \vec{p}}\right|_{\vec{R},\{n_A\},s}\;, \nonumber \\
T&=\frac{1}{\sqrt{g}}\left.\frac{\delta E}{\delta s}\right|_{\vec{p},\vec{R},\{n_A\}}\;, \nonumber \\
\vec{f}_{\rm rs}&=-\frac{1}{\sqrt{g}}\left.\frac{\delta E}{\delta \vec{R}}\right|_{\sqrt{g}\vec{p},\sqrt{g}s,\{\sqrt{g}n_A\}}
 \;, \nonumber \\
\mu^A&=\frac{1}{\sqrt{g}}\left.\frac{\delta E}{\delta n_A}\right|_{\vec{p},s,\vec{R},\{n_B|B\ne A\}}\;.
\end{align}

\subsection{A useful identity derived from reparametrization invariance}
An important element of the canonical thermodynamics of a capillary interface between two coexisting
fluids is the so-called Gibbs adsorption equation \cite{defayprigogine}, which relates together the
different intensive thermodynamic variables defining the state of the interface.  It results from
the fact that the excess thermodynamic free energies associated with such an interface scale
proportionally with the area of the interface at constant excess densities of the extensive quantities,
or equivalently, that the mechanical characterization of the interface is given by a single intensive
quantity, the surface tension. 

The thermodynamics of the type of membrane systems under our considerations is, however, different.
The canonical model of the membrane mechanics proposed by Helfrich and Evans provides an example.
In the model, the part of the free energy associated with the membrane mechanics is given by
$\int_\Sigma d A\;\left( 2\kappa H^2+\sigma_0\right)$, where $\kappa$ and $\sigma_0$ are constants.
It is clear that, if the linear size of the membrane surface is scaled by $\vec{R}\to \lambda\vec{R}$,
the term involving $\sigma_0$ will increase with $\lambda^2$ while the one including $\kappa$ (the bending
term) will not change.  Consequently, the Gibbs adsorption equation no longer exists; and the concept
of surface tension alone is no longer sufficient to describe the mechanical behaviour of the membrane
surface at mesoscopic or macroscopic scales.  Instead, as Eq.~(\ref{eq:varint}) implies, the restoring
force $\vec{f}_{\rm rs}$ provides an appropriate mechanical quantity.  

It is obvious from Eq.~(\ref{eq:varint}) that $\vec{f}_{\rm rs}$ depends on the determining (surface)
thermodynamic variables such as $\vec{v}$, $T$, $\mu^{A}$'s as well as the geometry of the dividing
surface.  It turns out that the tangential components of $\vec{f}_{\rm rs}$ are intimately connected
with the spatial inhomogeneities in $\vec{v}$, $T$, and $\mu^{A}$'s.  This connection arises from 
the fact that both the thermodynamic and the hydrodynamic behaviour of the dividing surface
is that of a two-dimensional ``fluid system."  In other words, they should remain invariant under
any change of the internal coordinate system.
  
To derive the explicit expression of the connection, we consider a situation where there exist spatial
inhomogeneities in the surface thermodynamic variables.  The total excess energy $E$ associated with
the dividing surface is then the surface integral of the local density $e$, i.e. a functional of $\vec{p}$,
$s$, $n_A$'s and $\vec{R}$.  Under an arbitrary, infinitesimal change of internal coordinates, or
``reparametrization" of the dividing surface,
\begin{equation}
{\uc'}^\alpha=\uc^\alpha+\delta \uc^\alpha\;,
\end{equation}
the functional form of the local density of excess entropy, as an example of a physical quantity, must
change from $s(\uc^1,\uc^2)$ in the old coordinate system to another form $s'({\uc'}^1,{\uc'}^2)$ in the
new such that 
\begin{equation}
s(\uc^1,\uc^2)=s'({\uc'}^1,{\uc'}^2)= s'(\uc^1,\uc^2)+\delta \uc^\alpha\partial_\alpha s\;,
\end{equation}
as required by the reparametrization invariance.  This is equivalent to making the following variation
in the functional form of the entropy density expressed in the old coordinate system: 
\begin{equation}
\label{eq:scalartrans}
\delta s \equiv s'(\uc^1,\uc^2)-s(\uc^1,\uc^2) = -\delta \uc^\alpha\partial_\alpha s\;.
\end{equation}
The variations in the functional forms of $\vec{p}$, $\vec{R}$ and $n_A$ can be expressed similarly.

Under fixed boundary conditions, the above variations lead to a variation in the integrated energy $E$
\begin{multline}
\label{E_integrated}
\delta E=\int_{\rm M} d^2\uc\;\bigg(\vec{v}\cdot\delta\left(\sqrt{g}\vec{p}\right)
         +T \delta\left(\sqrt{g}s\right)\\-\sqrt{g}\vec{f}_{\rm rs}\cdot\delta\vec{R}+\sum_A\mu^A\delta\left(\sqrt{g}n_A\right)\bigg)\ ,
\end{multline}   
where Eq.~(\ref{eq:varint}) has been used.  Obviously, this variation must be zero.

Inserting into Eq.~(\ref{E_integrated}) Eq.~(\ref{eq:scalartrans}), its analogs for $\vec{p}$, $n_A$'s and
$\vec{R}$, as well as
$\delta(\sqrt{g})=\sqrt{g}\vec{t}^\alpha\cdot\partial_\alpha\delta\vec{R}$, and performing a partial
integration yields
\begin{multline}
\delta E=\int_{\rm M} d A\;\bigg(\vec{p}\cdot\partial_\alpha\vec{v}+s\partial_\alpha T\\+\vec{f}_{\rm rs}\cdot\vec{t}_\alpha
         +\sum_A n_A\partial_\alpha\mu^A\bigg)\delta \uc^\alpha = 0 \;.
\end{multline}
Since $\delta \uc^\alpha$ is arbitrary, it can finally be concluded that 
\begin{equation}
\label{eq:GDrel}
\vec{f}_{\rm rs}\cdot\vec{t}_\alpha+\vec{p}\cdot\partial_\alpha\vec{v}+s\partial_\alpha T+\sum_A n_A\partial_\alpha\mu^A=0\;,
\end{equation}
must always hold.  This identity will become a useful one in the formulation of non-equilibrium
dynamics.  

Two remarks are worth making here.  First, although the physical reason underlying the above identity
appears conceptually obvious, we are not aware of any earlier work where the result has been systematically
derived.  Secondly, although we have made the derivation with membrane systems in mind, the result
also applies to a conventional capillary interface with a spatially varying surface tension $\sigma$, 
in which case $\vec{f}_{\rm rs}\cdot\vec{t}_\alpha = \partial_\alpha \sigma$.  Eq.~(\ref{eq:GDrel})
thus coincides with the expression of the Gibbs adsorption equation when it is applied to cases where
spatial inhomogeneities in the interface are present \cite{defayprigogine}.  

\section{Dynamic Formulation of Conservation Laws}
\label{sec:conslaws}
Having formulated a description of local thermodynamic equilibrium of the membrane-fluid system in terms
of the surface excess quantities, we now proceed to consider the general non-equilibrium state of the
system as defined in Introduction and develop a theory which describes the dynamics of the non-equilibrium
state.  Following the philosophy of developing an effective theory by the use of the Gibbs model system,
where two bulk fluids meet at an infinitely thin dividing surface, we assume that the hydrodynamic 
description of the two bulk fluids, in terms of their local thermodynamics and transport properties,
is entirely known, and thus we relate, not only the thermodynamic, but also the hydrodynamic, effects
associated with the microscopic inhomogeneity in the real system to the dividing surface and to its
modes of interaction with the two bulk fluids.

To make the presentation easy to follow, we first define a few notations pertaining to the description 
of the space.  Similar to previous notation, $\vec{R}(\uc^{1},\uc^{2},t)$ is used to represent the dynamic
shape of the dividing surface, and the space is then divided into two regions separated by the dividing
surface: ``$+$"-region refers to the bulk-fluid region which the normal of the surface,
$\vec{n}(\uc^{1},\uc^{2},t)$ points into and ``$-$"-region the other.  A scalar function $f(\vec{r},t)$
is introduced such that it is zero on the dividing surface and positive/negative on the $+$/$-$ side,
respectively; two Heaviside step functions are then defined as $\theta^{\pm}(\vec{r},t)=\theta(\pm f(\vec{r},t))$.
A few identities follow immediately,
\begin{align}
\label{math_identities}
& \partial_t\theta^\pm =\mp\int_{\rm M} d A\;\vec{n}\cdot\partial_t\vec{R}\;\delta\left(\vec{r}-\vec{R}\right)\;, \nonumber \\
& \vec{\nabla}\theta^\pm =\pm\int_{\rm M} d A\;\vec{n}\delta\left(\vec{r}-\vec{R}\right)\;, \nonumber \\
& \vec{t}_\alpha\cdot\vec{\nabla}\delta\left(\vec{r}-\vec{R}\right)
=-\partial_\alpha\delta\left(\vec{r}-\vec{R}\right) \;,
\end{align}
which will be used below.

The starting point of the hydrodynamic description is a formulation of the basic laws of
conservation of the molecular number of each species, momentum and energy for the model system
in the context of non-equilibrium dynamics.  Specifically, it is assumed that the following equation
of dynamics holds,
\begin{equation}
\label{eq:3dcons}
\frac{\partial}{\partial t}{\bar x}(\vec{r},t) = -\vec{\nabla}\cdot\vec{\bar{J}}_{X}(\vec{r},t) + \bar{\sigma}_{x}\;,
\end{equation}  
where ${\bar x}(\vec{r},t)$ represents the volume density of a conserved quantity $X$ and runs over
the number density of a species, ${\bar n}_{A}(\vec{r},t)$, the momentum density,
$\vec{\bar{p}}(\vec{r},t)$, and the energy density ${\bar e}(\vec{r},t)$, and where
$\vec{\bar{J}}_{X}(\vec{r},t)$ represents the corresponding flux.  This formulation is broad in that
it allows for the presence of a term $\bar{\sigma}_{x}$, which can account for ``sink-source" mechanisms in 
the dynamics of conserved quantities.  For example, when $X$ represent molecular numbers, $\bar{\sigma}_{x}$
can be used to describe the kinetics of chemical reactions.  In the case where an electric field
$\vec{E}$ is applied on the system, which may contain molecules carrying charges $\{ q^{A}\}$,
$\bar{\vec{\sigma}}_{p} = \vec{E}\sum_{A}q^{A}\bar{n}_{A}$ and 
$\bar{\sigma}_{e} = \sum_{A}q^{A}\bar{n}_{A} \vec{E}\cdot \vec{v}_{A}$ may be used to model the
effects of the electric field, where $\vec{v}_{A}$ denotes the velocity of species $A$. 

In principle, the law of angular-momentum conservation should also be included.  In standard hydrodynamic
descriptions of conventional fluids, the canonical approximation is that each local fluid element has no
intrinsic angular momentum.  In the case of the type of membrane-fluid systems under our considerations,
it would be expected that physical situations exist where the approximation is a valid one, and also that
in other situations it no longer holds.   But, we will work with the simpler cases where the approximation
may be made.  

Based on the structure of the model system, ${\bar x}(\vec{r},t)$ is expressed as 
\begin{multline}\label{model_def_x}
{\bar x}(\vec{r},t) =  {\bar x}^{+}(\vec{r},t)\theta^{+}(\vec{r},t) + {\bar x}^{-}(\vec{r},t)\theta^{-}(\vec{r},t)\\
+ \int_{\rm M} dA(\uc^{1},\uc^{2},t) \; x^{}(\uc^{1},\uc^{2},t) \;
                                           \delta\left(\vec{r} - \vec{R}(\uc^{1},\uc^{2},t)\right) \;,
\end{multline}
where ${\bar x}^{\pm}(\vec{r},t)$ represents the volume density of $X$ in the ``$\pm$"--bulk fluid, 
$\mathrm{M}$ indicates that the surface integral is over the whole dividing surface, and finally, 
$x(\uc^{1},\uc^{2},t)$ is the surface density of the excess of quantity $X$.  It follows
immediately from the concept of the model system and the assumption of local thermodynamic
equilibrium that ${\bar n_{A}}^{\pm}(\vec{r},t)$'s, $\vec{\bar{p}}^{\pm}(\vec{r},t)$,
${\bar e}^{\pm}(\vec{r},t)$ and ${\bar s}^{\pm}(\vec{r},t)$ satisfy the expressions of the bulk
thermodynamics, Eq.~(\ref{eq:firstbulk}) and \linebreak Eq.~(\ref{eq:GDbulk}).  Thus, the two equations provide
operational definitions of the corresponding chemical-potential fields, ${\bar \mu}^{A}_{\pm}(\vec{r},t)$'s,
hydrodynamic velocity fields $\vec{\bar{v}}^{\pm}(\vec{r},t)$, temperature fields,
${\bar T}^{\pm}(\vec{r},t)$, and pressure fields ${\bar p}^{\pm}(\vec{r},t)$.  Regarding
the thermodynamic characterization of the dividing surface, a similar assumption is made: the
non-equilibrium surface density of the excess energy, $e(\uc^{1},\uc^{2},t)$, is still functionally
related to the other surface densities, \linebreak $\vec{p}(\uc^{1},\uc^{2},t)$, $s(\uc^{1},\uc^{2},t)$, and
$n_{A}(\uc^{1},\uc^{2},t)$'s as well as the shape field $\vec{R}(\uc^{1},\uc^{2},t)$ according to
Eq.~(\ref{eq:varint}).  The conjugate variables defined thus by Eq.~(\ref{derivatives}),
$\vec{v}(\uc^{1},\uc^{2},t)$, \linebreak $T(\uc^{1},\uc^{2},t)$, and $\mu^{A}(\uc^{1},\uc^{2},t)$'s 
are then considered as the velocity, the temperature and the chemical potentials of the dividing
surface, respectively, and $\vec{f}_{\rm rs}(\uc^{1},\uc^{2},t)$ can be identified with the mechanical
force exerted on the dividing surface.  Consequently,
Eq.~(\ref{eq:peqofstate}) and Eq.~(\ref{eq:GDrel}) also hold.

Similar to that of ${\bar x}(\vec{r},t)$, the model expression of the corresponding flux
$\vec{\bar{J}}_{X}(\vec{r},t)$ is given by 
\begin{multline}
\label{model_def_flux}
\vec{\bar{J}}_X(\vec{r},t) = \vec{\bar{J}}^{+}_{X}(\vec{r},t) \theta^{+}(\vec{r},t)
                          + \vec{\bar{J}}^{-}_{X}(\vec{r},t) \theta^{-}(\vec{r},t)  \\
                          \shoveleft{\;\;+ \int_{\rm M} dA\;\Big[ \vec{j}_x^{(0)}(\uc^{1},\uc^{2},t)}\\
                            - \vec{j}_x^{(1)}(\uc^{1},\uc^{2},t)(\vec{n}\cdot\vec{\nabla})\Big]
                              \delta \left(\vec{r}-\vec{R}\right)\;,
\end{multline}
where $\vec{\bar{J}}^{\pm}_{X}(\vec{r},t)$ represents the flux in the ``$\pm$"--bulk fluid.  It is one of
the model statements that the functional dependence of $\vec{\bar{J}}^{\pm}_{X}(\vec{r},t)$ on the
relevant bulk hydrodynamic state variables and bulk transport coefficients is known.  Quantities 
$\vec{j}_x^{(0)}(\uc^{1},\uc^{2},t)$ and $\vec{j}_x^{(1)}(\uc^{1},\uc^{2},t)$ account for the
excess contributions due to the inhomogeneity in the real system.

The presence of a non-zero $\vec{j}_x^{(1)}(\uc^{1},\uc^{2},t)$ in the above model expression is
only necessary when $X$ represents linear momentum.  In that case $\vec{j}_{p}^{(1)}(\uc^{1},\uc^{2},t)$
can be identified as the surface flux of angular momentum. It
includes the contributions from both the motion and the mechanical stress in the material,
and is non-zero even when there is no motion.  The reason for it is that a complete
mechanical characterization of a membrane within the Gibbs model system requires both an effective
surface-stress tensor, included in $\vec{j}_{p}^{(0)}(\uc^{1},\uc^{2},t)$, and an effective,
non-zero bending-moment tensor, represented by a non-zero $\vec{j}_{p}^{(1)}(\uc^{1},\uc^{2},t)$, in
order that there is a mechanical equivalence between the real and the model systems when their
respective distributions of mechanical properties are integrated over the transverse dimension across
the inhomogeneous region.  A more detailed discussion of this issue will be given in Ref.~\cite{stress}.

Inserting the model expressions, Eq.~(\ref{model_def_x}) and Eq.~(\ref{model_def_flux}), in the
conservation law, Eq.~(\ref{eq:3dcons}), carrying out the partial differentiations using
Eq.~(\ref{math_identities}) and performing some partial integrations lead to the following
set of equations:
\begin{eqnarray}
\frac{\partial}{\partial t}{\bar x}^\pm & = &  -\vec{\nabla}\cdot \vec{\bar{J}}_X^\pm + \bar{\sigma}_{x}^{\pm} \;,
\label{bulk_consv} \\
D_t x^{} & = &  -D_\alpha \left[ \left(\vec{j}_x^{(0)} - x^{}\partial_t\vec{R} \right) \cdot\vec{t}^\alpha
                                + \vec{j}_x^{(1)} \cdot\vec{t}_\beta K^{\alpha\beta} \right] \nonumber \\
           &  &   - \left(\vec{J}_X^+\cdot\vec{n} - x^+\partial_t\vec{R}\cdot\vec{n}\right)\nonumber\\
&&            + \left(\vec{J}_X^-\cdot\vec{n} - x^-\partial_t\vec{R}\cdot\vec{n}\right) + \sigma_{x} \;, 
\label{surface_consv} \\
0 & = & - \left( \vec{j}_x^{(0)} - x^{}\partial_t\vec{R} \right) \cdot\vec{n} 
                 + D_\alpha \left(\vec{j}_x^{(1)} \cdot\vec{t}^\alpha \right) \;,
                  \label{eq:angular_moment} \\
0 & = & \vec{j}_x^{(1)} \cdot\vec{n}  
\label{eq:no_normal_bending}\;,
\end{eqnarray}
where a new differential operator with respect to $t$, \linebreak 
$D_t(\bullet) \equiv 1/\sqrt{g}\,\partial_t(\sqrt{g}\,\bullet)$, has been introduced.
$\vec{J}_{X}^{\pm}$ and $x^\pm$ represent the boundary values of the bulk hydrodynamic quantities,
i.e., the values of $\vec{\bar{J}}^\pm(\vec{r},t)$ and ${\bar x}^\pm(\vec{r},t)$ evaluated at the
dividing surface, respectively.  $\sigma_{x}$ is the surface excess of the rate of
generation/disappearance of quantity $X$ associated with the sink-source term in
Eq.~(\ref{eq:3dcons}).   

Eq.~(\ref{bulk_consv}) is a reiteration of one of the model statements that fluids with the
known bulk behavio\-urs fill the regions on the two sides of the infinitely thin dividing
surface.  Eq.~(\ref{surface_consv}) represents the set of equations governing the dynamics
of the relevant excess surface quantities.  Eq.~(\ref{eq:angular_moment}) and 
Eq.~(\ref{eq:no_normal_bending}) are simply conditions of self-consistency, implied in the
Gibbs-model construction, on the normal components of $\vec{j}_x^{(0)}$ and $\vec{j}_x^{(1)}$.

Based on Eq.~(\ref{surface_consv}), a tangential current within the dividing surface, denoted by 
$j^\alpha_x$, and two transverse currents entering/leaving the dividing surface, denoted by
$j^\pm_x $, can be identified:
\begin{align}
j^\alpha_x & \equiv ( \vec{j}_x^{(0)} - x\partial_t\vec{R} )\cdot\vec{t}^\alpha 
                 + \vec{j}_x^{(1)}\cdot\vec{t}_\beta K^{\alpha\beta} \;,\\
j^\pm_x & \equiv \vec{J}_{X}^{\pm}\cdot\vec{n}-x^\pm\partial_t\vec{R}\cdot\vec{n}\;. \label{transverse_j}
\end{align}
Eq.~(\ref{surface_consv}) can thus be written as
\begin{align}
D_t e& =-D_\alpha j^\alpha_e+j^{-}_e-j^{+}_e \;,
\label{eq:econs} \\
D_t \vec{p}& =-D_\alpha \vec{j}^\alpha_{p}+\vec{j}^{-}_{p}-\vec{j}^{+}_{p}  \;,
\label{eq:momcons} \\
D_t n_A& =-D_\alpha j^\alpha_A + j^{-}_A-j^{+}_A + \sum_K \nu_{A,K}\xi^K \;, 
\label{eq:continuity}
\end{align}
for cases where $\sigma_{p} = \sigma_{e} = 0$.  We will only consider such cases in what follows.

The last term in Eq.~(\ref{eq:continuity}) has been added to allow for the possibility of ``chemical 
reactions" taking place in the membrane.  The summation index $K$ runs over all possible reactions,
$\xi^K$ is the rate of reaction $K$ per unit area and $\nu_{A,K}$ the stoichiometric coefficient of
species $A$ in reaction $K$.  The word ``chemical reaction" in our theory should be understood in
a broader sense than that pertaining to a genuine chemical reaction.  Connected to the assumption
used in the formulation of thermodynamics that the membrane is considered to be quasi-statically
impermeable to certain chemical species, non-equilibrium transport across the membrane of any of
those chemical species (denoted by $\bar{C}$), such as the flip-flop process of a particular lipid
species, is modelled by the following ``reaction," 
\begin{equation}
\label{eq:flipflop_1}
\bar{C}^{+} \stackrel{\displaystyle \longrightarrow}{\longleftarrow} \bar{C}^{-} \;,
\end{equation}
where $\bar{C}^{+}$ and $\bar{C}^{-}$ are considered as two different labelled species.

It must have not escaped the reader that Eq.~(\ref{model_def_x}) and Eq.~(\ref{model_def_flux})
on their own are not sufficient to define $x(\uc^{1},\uc^{2},t)$, $\vec{j}_x^{(0)}(\uc^{1},\uc^{2},t)$
and $\vec{j}_x^{(1)}(\uc^{1},\uc^{2},t)$.  At the conceptual level, the definitions of the 
surface excess quantities may be understood in the following sense.  Consider the cell 
${\bar{\Sigma}}$ defined in Section 3 and bear in mind in particular that $\epsilon^{\pm}$ 
must be chosen such that the hydrodynamic behaviour of the real system coincides with that
of the model system outside the top and bottom surfaces of the cell.  If $X_{\bar{\Sigma}}$
denotes the amount of quantity $X$ in the cell in the real system, the following condition
of equivalence between the real and the model systems 
\begin{equation}
\label{matching_x}
\int_{\bar{\Sigma}} dV\;{\bar x}(\vec{r},t) = X_{\bar{\Sigma}} 
\end{equation}
then defines $x(\uc^{1},\uc^{2},t)$.   In a similar fashion, a number of conditions of equivalence
should be satisfied by the model quantities $\vec{j}_x^{(0)}(\uc^{1},\uc^{2},t)$ and
$\vec{j}_x^{(1)}(\uc^{1},\uc^{2},t)$.  If $\bar{B}^{\alpha}$ represents the cross section of
${\bar{\Sigma}}$ at constant $\uc^{\alpha}$, then the following model quantity
\begin{equation}
\label{matching_current}
\int_{\bar{B}^\alpha} d{\vec{\bar{A}}}\cdot\left[ \vec{\bar{J}}_X-\bar{x}\partial_t\left(\vec{R}
     +h\vec{n}\right)\right] \;,
\end{equation}
where $d{\vec{\bar{A}}}$ is a normally directed area element on $\bar{B}^{\alpha}$ and where the
integration is taken over the transverse dimension, $h$, along the normal $\vec{n}$, should equal
to the total current crossing $\bar{B}^{\alpha}$ in the real system.  

The discussion in the preceding paragraph in fact gives rise to a subtle issue concerning the
relationship between the surface density of the excess momentum, $\vec{p}$, defined by Eq.~(\ref{matching_x}),
and the excess current associated with the total mass, $\vec{j}_{\rho}^{(0)}$.  Given $\vec{p}$,
the hydrodynamic velocity $\vec{v}$ associated with the dividing surface is determined by the local
thermodynamics, i.e., 
\begin{equation}
\vec{v} = \vec{p}/\rho \;,
\end{equation}
where $\rho$ is the surface density of the excess mass.  $\vec{j}_{\rho}^{(0)}$, defined by the
condition Eq.~(\ref{matching_current}), is not equal to $\vec{p}$ in principle.
\footnote{ This is due to the fact that, when the dividing surface is curved, the volume measure and
the area measure will then have a different dependence on the distance to the dividing surface.}
The difference is on the order of $\epsilon^{\pm}/R$ where $R$ is curvature, and it will be neglected,
to a first approximation.  In what follows, we will thus use 
\begin{equation}
\label{mass_current}
\vec{j}_{\rho}^{(0)} = \rho \vec{v} \;.
\end{equation}
By this approximation, Eq.~(\ref{eq:angular_moment}) reduces to 
\begin{equation}
\label{v_normal}
\vec{v}\cdot\vec{n}=\partial_t\vec{R}\cdot\vec{n}\;,
\end{equation}
a description consistent with our intuition.  An alternative expression of Eq.~(\ref{v_normal}),
which will also be used later, is
\begin{equation}
\label{surface_v}
\vec{v} = v^{\alpha}\vec{t}_{\alpha} + \partial_t\vec{R} \;,
\end{equation}
where $v^{\alpha}\vec{t}_{\alpha}$ can be interpreted as the {\em surface-intrinsic} part of $\vec{v}$. 

\section{Entropy production}
\label{sec:eprod}
The set of equations of dynamics derived in the previous section, Eq.~(\ref{eq:econs}),
Eq.~(\ref{eq:momcons}) and Eq.~(\ref{eq:continuity}), involve both the currents describing the
transport processes in the tangent space of the dividing surface, $j_{x}^{\alpha}$, and the
currents describing the transverse transport processes, which in turn involve the boun\-dary
values of the bulk hydrodynamic fields.  In order that the equations form a closed set, 
relations between the currents and the surface thermodynamic/hydrodynamic fields should be
developed from the equation of entropy balance, according to one of the basic principles of
non-equilibrium thermodynamics.  In this section, we present the derivation of the equation
of entropy balance from which entropy production associated with transport processes
is identified \cite{prigogine}. 

The tangential currents can be decomposed into two parts, a reactive part, which is associated
with reversible transport processes, and a dissipative part, which is associated with irreversible
processes:
\begin{align}
j^\alpha_A &= j^\alpha_{A,{\rm r}} + j^\alpha_{A,{\rm d}}\;,\\ 
\vec{j}^\alpha_{p} &= \vec{j}^\alpha_{p,{\rm r}} + \vec{j}^\alpha_{p,{\rm d}}\;,\\
j^\alpha_e &= j^\alpha_{e,{\rm r}} + j^\alpha_{e,{\rm d}}\;,
\end{align}
where the subscript ``${\rm r}$'' denotes the reactive part and ``${\rm d}$'' the dissipative.
Once an expression for the entropy production is obtained, the reactive parts of the currents
can be determined, by the definition that they should not contribute to the entropy production;
moreover, and more importantly, thermodynamic/hydrodynamic ``forces" driving the dissipative
currents can also be identified \cite{chaikin,reichl}.  

The equation of entropy balance in its general form can be written as
\begin{equation}
\label{general_s}
D_t s=-D_\alpha j^\alpha_s + j_s^- - j_s^+ + \sigma_s \;,
\end{equation}
where $j^\alpha_s$ represents the tangential current of entropy, $j_s^\pm$ are transverse
currents, and $\sigma_s$ is the density of entropy production.  The derivation of an explicit
expression of the equation starts from calculating, based on Eq.~(\ref{eq:varint}), the variations
with time of all the thermodynamic quantities and isolating $D_t s$ as
\begin{multline}
\label{eq:sprod}
D_t s=\frac{1}{T}\Big[D_t e+\vec{f}_{\rm rs}\cdot\partial_t \vec{R}-\vec{v}\cdot D_t\vec{p}\\-\sum_A\mu^A D_t n_A
      -D_\alpha\left(\vec{S}^\alpha\cdot\partial_t\vec{R}\right)\Big] \;.
\end{multline}
Further derivation can be carried out by using the conservation laws, Eq.~(\ref{eq:econs})
to Eq.~(\ref{eq:continuity}), to replace the corresponding time derivatives and using the identity
derived in Section 3.2, Eq.~(\ref{eq:GDrel}), to replace $\vec{f}_{\rm rs}\cdot\partial_t \vec{R}$ 
with 
\begin{multline}
\vec{f}_{\rm rs}\cdot\partial_t \vec{R}=\vec{f}_{\rm rs}\cdot\left(\vec{v}-v^\alpha\vec{t}_\alpha\right)\\
                              =\vec{f}_{\rm rs}\cdot\vec{v}
                               +v^\alpha \left( \vec{p}\cdot\partial_\alpha\vec{v}+s\partial_\alpha T
                                              +\sum_A n_A\partial_\alpha\mu^A \right)\;,
\end{multline}
where Eq.~(\ref{surface_v}) has been used.  With the use of $\vec{f}_{\rm rs}=D_\alpha\vec{S}^\alpha_{(0)}$
in addition, the equation of entropy balance can be rearranged into
\begin{align}
\label{s_production}
D_t s=&-D_\alpha\bigg[\frac{1}{T} \bigg(j^\alpha_e+\vec{S}^\alpha\cdot\partial_t\vec{R}-\sum_A\mu^A j^\alpha_A \nonumber\\
&\phantom{-D_\alpha\bigg[\frac{1}{T} \bigg(}-\vec{v}\cdot\vec{j}^\alpha_{p}-\vec{v}\cdot\vec{S}^\alpha_{(0)}\bigg)\bigg] 
 + j_s^- - j_s^+ \nonumber\\
&+ \left[ \bigg( j^\alpha_e+\vec{S}^\alpha\cdot\partial_t\vec{R}
              - \sum_A\mu^A j^\alpha_A-\vec{v}\cdot\vec{j}^{\alpha}_{p} \right. \nonumber\\
&\phantom{+\bigg(}-\vec{v}\cdot\vec{S}^\alpha_{(0)}-T s v^\alpha \bigg) \partial_\alpha\frac{1}{T} \nonumber\\
&- \frac{1}{T} \left( \vec{j}^\alpha_{p}+\vec{S}^\alpha_{(0)}-\vec{p} v^\alpha \right)
                  \cdot\partial_\alpha \vec{v} \nonumber\\
& \left. - \frac{1}{T}\sum_A \left( j^\alpha_A-n_A v^\alpha \right)
                  \partial_\alpha\mu^A - \frac{1}{T}\sum_K \xi^K \Gamma_K \right]\nonumber\\
& + \left[ \frac{1}{T} \left( j^-_e-j^+_e \right)
       -\frac{1}{T}\vec{v}\cdot \left( \vec{j}_{p}^--\vec{j}_{p}^+ \right) \right. \nonumber\\
& \left. -\frac{1}{T}\sum_A\mu^A \left( j^-_A-j^+_A \right) - (j_s^- - j_s^+) \right] \;,
\end{align}
where $\Gamma_K \equiv \sum_A\mu^A\nu_{A,K}$ is the affinity for reaction $K$.  The interpretations
of the different terms in the equation are made clear immediately by a comparison with the
general form, Eq.~(\ref{general_s}).  The terms collected in the first pair of square brackets give 
the tangential current of entropy, 
\begin{align}
\label{j_s}
j_{s}^{\alpha} = \, \frac{1}{T} \bigg(& j^\alpha_e + \vec{S}^\alpha\cdot\partial_t\vec{R} - \sum_A\mu^A j^\alpha_A
                            -\vec{v}\cdot\vec{j}^\alpha_{p}\nonumber\\
&-\vec{v}\cdot\vec{S}^\alpha_{(0)}\bigg)
                            \;.
\end{align}
The terms in the second and third pairs of square brackets sum up the entropy production from all of the
membrane-related transport processes.  Those in the second pair represent the entropy production
from transport processes intrinsic to the dividing surface, and those in the third describe the entropy
production from processes of transport between the bulk fluids and the dividing surface.

\subsection{The intrinsic transport processes} 
The reactive parts of the currents can now be determined from the explicit expression of the
entropy production given in Eq.~(\ref{s_production}).  An examination of the terms enclosed by
the second pair of square brackets yields
\begin{equation}
\label{eq:DTrfrs}
\vec{j}^\alpha_{p,{\rm r}} = -\vec{S}^\alpha_{(0)} + \vec{p} v^\alpha\;,
\end{equation}
\begin{equation}
\label{A_current_r}
j^\alpha_{A,{\rm r}} = n_A v^\alpha \;,
\end{equation}
and 
\begin{equation}
\label{eq:jer}
j^\alpha_{e,{\rm r}} = v^\alpha\left(\vec{v}\cdot\vec{p}+Ts+\sum_A\mu^A
n_A\right)-\vec{S}^\alpha\cdot\partial_t\vec{R}\;,
\end{equation}
because these currents alone, in the absence of the dissipative parts, make no contribution to the
entropy production.  Finally, it can be concluded that the reactive and dissipative parts
of $j_{s}^{\alpha}$ defined in Eq.(\ref{j_s}) are given, respectively, by
\begin{align}
j^\alpha_{s,{\rm r}}& = s v^\alpha\;,\\
j^\alpha_{s,{\rm d}}& = \frac{1}{T} \left( j^\alpha_{e,{\rm d}} - \vec{v}\cdot\vec{j}^\alpha_{p,{\rm d}}
                       -\sum_A\mu^A j^\alpha_{A,{\rm d}}\right) \;. \label{eq:jsd}
\end{align}
In the most general sense, the above identifications of the reactive currents are not complete.
We will discuss this point again where constitutive relations are derived.

The various dissipative currents, $\vec{j}^\alpha_{p,{\rm d}}$, $j^\alpha_{A,{\rm d}}$'s,
and $j^\alpha_{e,{\rm d}}$ can now be determined.   However, not all of the $j^\alpha_{A,{\rm d}}$'s
are independent due to a constraint 
\begin{equation}
\label{eq:ddep}
\sum_A m^A {j}^\alpha_{A,{\rm d}}=0  \;,
\end{equation}
which follows from Eq.~(\ref{mass_current}) and Eq.~(\ref{A_current_r}).  
Thus, $j^\alpha_{A,{\rm d}}$ of any species $A$ can be chosen to be the one dependent on the
rest.  Given that a judicious choice $A=O$ can be made for one reason or another, Eq.~(\ref{eq:ddep})
can be written as
\begin{equation}
\label{eq:j1}
{j}^\alpha_{\w,\rm d} = -\sum_{A \ne \w} \frac{m^A}{m^{\w}} {j}^\alpha_{A,{\rm d}}\;.
\end{equation}

The entropy production from the intrinsic dissipative processes can finally be expressed in 
terms of the various independent dissipative currents, $\vec{j}^\alpha_{p,{\rm d}}$,
$j^\alpha_{A\ne \w,{\rm d}}$'s, and $j^\alpha_{e,{\rm d}}$ and the thermodynamic forces driving them: 
\begin{align}
T\sigma_{s,\parallel} = & -{j}^\alpha_{s,{\rm d}}\partial_\alpha T
                       - \vec{j}^\alpha_{p,{\rm d}}\cdot\partial_\alpha\vec{v}
                       - \sum_{A\ne \w} {j}^\alpha_{A,{\rm d}} \partial_\alpha \tilde{\mu}^A\nonumber\\
&                       - \sum_K \xi^K \Gamma_K \;,\label{eq:s_production_tang}
\end{align}
where 
\begin{equation}
\tilde{\mu}^A \equiv \mu^A - \frac{m^A}{m^{\w}}\mu^{\w}
= \frac{1}{\sqrt{g}} \left.\frac{\delta E}{\delta n_A}\right|_{\vec{p},\rho,s,\vec{R},\{n_{B\ne A,\w} \}} \;.
\end{equation}

\subsection{The processes of transport between the surface and the bulk fluids}
The processes of transport between the surface and the bulk fluids contribute to the total entropy
production in the form of those terms contained in the last pair of square brackets in Eq.~(\ref{s_production}).
It is clear that the contributions depend not only on the boundary behaviour of the bulk hydrodynamic
fields, but also on the surface hydrodynamic fields.  A more illuminating expression of the contributions
can be derived as follows.

The values of the bulk currents evaluated at the dividing surface appear in Eq.~(\ref{transverse_j}) which
defines the transverse currents, $j_{e}^{\pm}$, $\vec{j}_{p}^{\pm}$, $j_{A}^{\pm}$'s and $j_{s}^{\pm}$,
and they are given, respectively, by
\begin{equation}
\vec{J}_{e}^{\pm} = (e^{\pm} + p^{\pm})\vec{v}_{\pm} - \mathsf{T}_{\rm d}^\pm \cdot \vec{v}_{\pm}
                 +\vec{J}_{q}^{\pm} \;,
\end{equation}
where $\mathsf{T}_{\rm d}^\pm$ is the viscous stress of the corresponding bulk fluid and
$\vec{J}_{q}^{\pm}$ is the heat flux; 
\begin{equation}
\mathsf{J}^\pm_{p} = \rho^\pm \vec{v}_\pm\vec{v}_\pm + p^\pm \mathsf{I}-\mathsf{T}_{\rm d}^\pm\;,
\end{equation} 
where $\mathsf{I}$ is the identity tensor; 
\begin{equation}
\vec{J}_{A}^{\pm} = n_{A}^{\pm} \vec{v}_{\pm} + \vec{J}_{A,{\rm d}}^{\pm}  \;;
\end{equation} 
and 
\begin{equation}
\vec{J}_{s,\rm{d}}^{\pm} = (\vec{J}_{q}^{\pm} - \mu^{A}_{\pm} \vec{J}_{A,{\rm d}}^{\pm})/T_{\pm}  \;.
\end{equation} 
Inserting these explicit expressions into Eq.~(\ref{transverse_j}) yields 
\begin{align}
\label{eq:momext}
\vec{j}_{p}^\pm   = & \; \rho^\pm \left[ \vec{n}\cdot \left(\vec{v}_\pm - \vec{v}\right) \vec{v}_\pm \right]
                  + p^\pm\vec{n} -\vec{n}\cdot \mathsf{T}_{\rm d}^\pm \;, \\
j^\pm_e  = & \; \rho^{\pm}\vec{v}_\pm^2 \left[ \vec{n} \cdot\left(\vec{v}_\pm-\vec{v}\right) \right]
              + p^\pm \vec{n} \cdot \vec{v}\nonumber\\&   - \vec{n} \cdot \mathsf{T}^\pm_{\rm d}\cdot\vec{v}_\pm 
            + \sum_{A} \mu^{A}_\pm j_{A}^{\pm} + T_\pm j_{s}^{\pm} \;,
\end{align}
where $\vec{n} \cdot \vec{v} = \vec{n} \cdot \partial_t\vec{R}$ has been used.

The sum of the entropy production from all the processes of transport between the surface and the bulk
fluids can now be expressed as 
\begin{eqnarray}
\label{s_production_perp_1}
T\sigma_{s,\perp} & \equiv & j^-_e - j^+_e - \vec{v}\cdot\left( \vec{j}_{p}^- -\vec{j}_{p}^+ \right)\nonumber\\
&&                        -\sum_A\mu^A\left(j^-_A-j^+_A\right)-T \left( j^-_s - j^+_s \right)  \nonumber \\
               & = & \sum_{ A} \left[ {j}_{A}^{-}\left(\mu^{A}_--\mu^A\right) - {j}_{A}^{+} \left(\mu^{A}_+-\mu^A\right) \right]\nonumber\\
&&                    + j^-_s \left( T_- - T \right) - j^+_s \left(T_+ - T\right) \nonumber \\
               &   & + \bigg( -\vec{n}\cdot\mathsf{T}^-_{\rm d}\cdot\vec{n} + \frac{1}{2}\rho^- \left(\vec{v}_{-}-\vec{v}\right)^2\nonumber\\
&&\phantom{+\bigg(}                    + \frac{1}{2} \rho^{-}\vec{v}^2_{-} - \frac{1}{2}\rho^{-}\vec{v}^2 \bigg) 
                      \vec{n}\cdot\left(\vec{v}_- - \vec{v}\right) \nonumber \\
               &   & + \bigg( \vec{n}\cdot\mathsf{T}^+_{\rm d}\cdot\vec{n} - \frac{1}{2}\rho^+ \left(\vec{v}_{+}-\vec{v}\right)^2\nonumber\\
&&\phantom{+\bigg(}
                    - \frac{1}{2} \rho^{+}\vec{v}^2_{+} + \frac{1}{2}\rho^{+} \vec{v}^2 \bigg) 
                      \vec{n}\cdot\left(\vec{v}_+ - \vec{v}\right) \nonumber \\ 
               &   & -\left( \vec{n}\cdot\mathsf{T}^-_{\rm d}\cdot\vec{t}_\alpha \right)
                      \left[ \; \vec{t}^\alpha\cdot \left( \vec{v}_--\vec{v} \right) \right]\nonumber\\
&&+ \left( \vec{n}\cdot\mathsf{T}^+_{\rm d}\cdot\vec{t}_\alpha \right)
                      \left[ \, \vec{t}^\alpha\cdot \left( \vec{v}_+-\vec{v} \right) \right] \;.
\end{eqnarray}
In the above expression, the currents $j_{A}^{\pm}$'s and $\vec{n}\cdot\left(\vec{v}_{\pm}- \vec{v}\right)$
are related by 
\begin{equation}
\sum_A m^A {j}^\pm_{A}=\rho^\pm \vec{n}\cdot \left( \vec{v}_\pm - \vec{v} \right) \;,
\end{equation}
which follows from $\sum_A m^A \vec{J}_{A}^{\pm} = \rho^{\pm} \vec{v}^{\pm}$, Eq.~(\ref{transverse_j})
and Eq.~(\ref{v_normal}).
Consequently, ${j}^\pm_{A}$'s for all $A$ values may be \linebreak chosen as the independent currents, or
alternatively, \linebreak $\rho^\pm \vec{n}\cdot \left( \vec{v}_\pm - \vec{v} \right)$ and
$\{ j^\pm_{A}, A \ne 0^{\pm} \}$ may be used as the independent currents, where $0^+$ and $0^-$
denote the two judiciously chosen species, whose currents will be eliminated explicitly.  

Making the latter choice and rewriting Eq.~(\ref{s_production_perp_1}) finally gives 
\begin{eqnarray}
\label{s_production_perp_2}
T\sigma_{s,\perp} & = &  {j}^-_s \left(T_- - T\right) - {j}^+_s \left(T_+ - T\right) \nonumber\\
                &   & + \sum_{A\ne 0^\pm} {j}_{A}^{-} \Delta\mu^{A}_{-} - \sum_{A\ne 0^\pm} {j}_{A}^{+} \Delta\mu^{A}_{+} 
                       \nonumber \\
               &  & + \left(-\vec{n}\cdot\mathsf{T}^-_{\rm d}\cdot\vec{n} + \Pi_-\right) \vec{n}\cdot\left(\vec{v}_--\vec{v}\right)\nonumber\\
&&                   +\left(\vec{n}\cdot\mathsf{T}^+_{\rm d}\cdot\vec{n} - \Pi_+\right) \vec{n}\cdot\left(\vec{v}_+-\vec{v}\right)\nonumber\\
               &   & -\left( \vec{n}\cdot\mathsf{T}^-_{\rm d}\cdot\vec{t}_\alpha \right)
                      \left[\, \vec{t}^\alpha\cdot \left( \vec{v}_--\vec{v} \right) \right]\nonumber\\
&&                    + \left( \vec{n}\cdot\mathsf{T}^+_{\rm d}\cdot\vec{t}_\alpha \right)
                      \left[\, \vec{t}^\alpha\cdot \left( \vec{v}_+-\vec{v} \right) \right] \;.
\end{eqnarray}
The newly-introduced quantities, $\Delta\mu^{A}_{\pm}$ and $\Pi_{\pm}$, are defined by 
\begin{align}
\Delta\mu^A_{\pm}  \equiv& \left(\mu^A_{\pm}-\frac{m^A}{m^{0^\pm}}\mu^{0^\pm}_{\pm}\right)\nonumber\\
&                      -\left(\mu^A -\frac{m^A}{m^{0^\pm}}\mu^{0^\pm}\right)\;, \quad   A\ne 0^\pm \;,  \\
\Pi_\pm  \equiv &\;\frac{\rho^\pm}{m^{0^\pm}} \bigg[ \frac{1}{2}m^{0^\pm} \left(\vec{v}_\pm-\vec{v}\right)^2
             + (\mu^{0^\pm}_\pm + \frac{1}{2}m^{0^\pm}\vec{v}^2_\pm )\nonumber\\
&\phantom{\frac{\rho^\pm}{m^{0^\pm}} \bigg[}    - (\mu^{0^\pm} + \frac{1}{2}m^{0^\pm}\vec{v}^2) \bigg] \;.
\end{align}
Their physical interpretations become more evident when it is recalled that
$(\mu^{A}_\pm + \frac{1}{2}m^{A}\vec{v}^2_\pm )$ and $(\mu^{A} + \frac{1}{2}m^{A}\vec{v}^2 )$ are,
respectively, the bulk and surface chemical potentials of species $A$ in the corresponding rest frames.  

\section{Constitutive relations}
\label{sec:consrel}
The expressions of the entropy production derived in the previous Section allow us to identify
the conjugate ``force"-\linebreak current pairs associated with all the different dissipative processes.
In this Section, we describe how physically meaningful relations between the currents and the forces
are developed, under the assumption that the non-equilibrium dynamics of the system is within the
linear regime.  To follow the standard terminology of statistical physics, we will call those relations
constitutive relations in general \cite{onsager,casimir45,kreuzer}.  

\subsection{Symmetry based classification of forces and currents}
Eq.~(\ref{eq:s_production_tang}) and Eq.~(\ref{s_production_perp_2}) involve many different currents,
such as ${j}^\alpha_{s,{\rm d}}$, $\vec{j}^\alpha_{p,{\rm d}}$, ${j}^\alpha_{A,{\rm d}}$'s, etc.,
and many different driving forces, such as $\partial_\alpha T$, $\partial_\alpha\vec{v}$,
$\partial_\alpha \tilde{\mu}^A$, etc..\footnote{In the formalistic sense what quantities are called forces
and what are called currents are completely arbitrary.  The convention we have adopted here conforms
with either physical intuition or historical usages.}  It is well known that a single current may be
driven by several different forces.   A systematic way to identify all the possible different forces
driving a particular current is the following: first, to classify all the forces and currents according
to their behaviour under a group of orthogonal transformations, consisting of both rotations of the
internal coordinate system and the inversion of the local normal vector to the dividing surface; and then,
to determine whether the symmetry of the system allows for, or forbids, a force of one
type, generically represented by $F^{i}$, to drive a current of another type, denoted $J_{j}$.  Here,
the Roman superscripts/subscripts label such classifications of forces and currents.

The generic classes of behaviour of a quantity under the coordinate transformations consist of
the following: scalar, vector, and tensors of different ranks, which transform like a scalar, vector,
and tensor with respect to internal coordinate transformations, but which remain invariant with
respect to an inversion of $\vec{n}$; pseudo-scalar, pseudo-vector, and pseudo-tensors, which are
different from scalar, vector, and tensors only in that they change their sign under the inversion
of $\vec{n}$.  

For the intrinsic dissipative processes involved in \linebreak Eq.~(\ref{eq:s_production_tang}), the
classification with respect to internal coordinate transformations is more essential.  When applied
this leads to the following conclusions: 
\begin{itemize}
\item[a)] {\em Genuine} chemical reactions involve forces $F^{K}$ and currents $J_{K}$, which
are given, respectively, by
\begin{equation}
F^{K} = - \Gamma_{K} = -\sum_{A} \mu^{A}\nu_{A,K} \;, \quad \quad    J_{K} = \xi^{K}  \;.
\end{equation}
Whether $F^{K}$ and $J_{K}$ are scalars or pseudoscalars depends on the precise nature of a reaction.
However, when a chemical reaction process $K^{*}$ refers to the ``flip-flop" from one side of the membrane
to the other of molecules of a particular chemical species, as described by Eq.~(\ref{eq:flipflop_1}),
the corresponding force and current 
\begin{equation}
\label{eq:flipflop_2}
F^{K^{*}} = - (\mu^{\bar{C}^+} - \mu^{\bar{C}^-} ) \;,
\quad \quad J_{K^{*}} = \xi^{K^*} \;,
\end{equation}
are clearly pseudoscalars. 
\item[b)]  Heat conduction and diffusion in the surface involve forces, $F^{s}_\alpha$ and 
$\{ F^{A}_\alpha , A\ne \w \}$, and currents, ${j}^\alpha_{s,{\rm d}}$ and
$\{ j^\alpha_{A,{\rm d}}, A\ne \w \}$,
which are
\begin{eqnarray}
\label{heat_diffusion}
F^{s}_\alpha  &\equiv -\partial_\alpha T \;,\quad \quad \quad 
                   & F^{A}_\alpha \equiv - \partial_\alpha \tilde{\mu}^A \; , \nonumber \\
J_{s}^\alpha  &\equiv {j}^\alpha_{s,{\rm d}} \;, \quad \quad \quad \quad  & J_{A}^\alpha\equiv j^\alpha_{A,{\rm d}} \;.    
\end{eqnarray}
These forces and currents transform like vectors under the rotations of the internal coordinate system.
\item[c)] The dissipative momentum transport involves forces and currents, whose representations need both
a tensor and a vector, 
\begin{align}
\label{surface_viscosity}
\partial_\alpha\vec{v} &= \left( \partial_\alpha\vec{v} \cdot \vec{t}_{\beta} \right) \vec{t}^{\beta}
                    + \left( \partial_\alpha\vec{v} \cdot \vec{n} \right) \vec{n} \nonumber\\
&                  \equiv -( F_{\alpha \beta} \vec{t}^{\beta} + F_{(n),\alpha} \vec{n} )\;,  \nonumber  \\
\vec{j}^\alpha_{p,{\rm d}}& = \left( \vec{j}^\alpha_{p,{\rm d}} \cdot \vec{t}^{\beta} \right) \vec{t}_{\beta}
                          + \left( \vec{j}^\alpha_{p,{\rm d}} \cdot \vec{n} \right) \vec{n} \nonumber\\
&                       \equiv J^{\alpha \beta}_{p} \vec{t}_{\beta} + J_{p,(n)}^{\alpha} \vec{n} \;.
\end{align}
Thus, $F_{(n),\alpha}$ and $J_{p,(n)}^{\alpha}$ form the (pseudo)vector force-current pair.  The
tensorial part, $F_{\alpha \beta}$, is not a symmetric tensor in general and can be decomposed into
three contributions, 
\begin{align}
\label{F_tensor}
F_{\alpha \beta} = \frac{1}{2} \Big[& g_{\alpha \beta} F_{\phantom{0}\gamma}^{\gamma}
                 + \left( F_{\alpha \beta} + F_{\beta \alpha} 
                        -g_{\alpha\beta} F_{\phantom{0}\gamma}^{\gamma} \right)\nonumber\\
&                 + \varepsilon_{\alpha \beta} \varepsilon^{\gamma \delta}F_{\gamma \delta} \Big] \;,
\end{align}
each of which is invariant under any internal-coordinate transformation and each of which should
be considered as an independent force.  The corresponding currents in $J_{\alpha \beta}$ can be
identified in a similar way.  
\end{itemize}

For the transport processes that give rise to the entropy production given in Eq.~(\ref{s_production_perp_2}),
it is more meaningful to use linear combinations of the apparent forces and currents to generate
new forces and currents that either remain invariant or change sign under an inversion of $\vec{n}$.
Before we discuss the new forces and currents, a change of notation will be made concerning our
reference to species.  So far, the definition of the labelled species, represented by $A$, is used
to keep the derivation concise.  In what follows, the chemical identities of the molecular species
will be explicitly referred to, in order that the  physical interpretations of quantities associated
with material transport become more apparent.  Specifically, $C$ will be used to represent those
molecular species, which have been assumed to reach chemical equilibrium across the membrane
quasi-statically, while $\bar{C}$ will denote those molecular species \linebreak which are assumed not to
permeate the membrane quasi-\linebreak statically.  With this notation and with the choice that the
``0" in the species expression $0^{\pm}$ used in Eq.~(\ref{s_production_perp_2}) refers to one of the
$\bar{C}$ species, the relevant terms in Eq.~(\ref{s_production_perp_2}) acquire an alternative
expression
\begin{multline}
\sum_{A\ne 0^{\pm}} {j}_{A}^{-} \Delta\mu^{A}_{-} - \sum_{A\ne 0^{\pm}} {j}_{A}^{+} \Delta\mu^{A}_{+}
\\\shoveleft{\;\;\;\;= \sum_{C} \left( {j}_{C}^{-} \Delta\mu^{C}_{-} - {j}_{C}^{+} \Delta\mu^{C}_{+} \right)}\\
  + \sum_{\bar{C}\ne 0} \left( {j}_{\bar{C}^{-}}^{-} \Delta\mu^{\bar{C}^{-}}_{-}  
                           -  {j}_{\bar{C}^{+}}^{+} \Delta\mu^{\bar{C}^{+}}_{+} \right) \; .
\end{multline}

From Eq.~(\ref{s_production_perp_2}) the new forces and currents can now be derived for the various different
transport processes, as follows. 
\begin{itemize}
\item[a)] The first four lines of the equation sum up the contributions from the absorption and
conduction of heat by the surface as well as from the adsorption onto, and permeation across, the surface
of the constituent molecules.  Genuine scalar forces $F_{({\rm s})}$ and currents $J^{({\rm s})}$, as well
as pseudoscalar forces $F_{({\rm a})}$ and currents $J^{({\rm a})}$, are given by
\begin{align}
 F^{s}_{({\rm s})}  &= -\left( T^+ + T^- -2T \right) \;, \nonumber \\
 J^{({\rm s})}_{s}  &= \frac{1}{2}\left( j_{s}^+ - j_{s}^- \right) \;, \\
 F^{s}_{({\rm a})}  &= -\left( T^+ - T^- \right) \;, \nonumber \\
 J^{({\rm a})}_{s}  &= \frac{1}{2}\left( j_{s}^+ + j_{s}^- \right) \;,
\end{align}
\begin{align}
F^{C}_{({\rm s})} &= -\left( \Delta\mu^{C}_{+} + \Delta\mu^C_{-} \right) \;, \nonumber \\
J_{C}^{({\rm s})} &= \frac{1}{2}\left( j^+_C - j^-_C \right) \;, \\
F^{C}_{({\rm a})} &= -\left( \Delta\mu^{C}_{+} - \Delta\mu^C_{-} \right) \;, \nonumber \\
J_{C}^{({\rm a})} &= \frac{1}{2}\left( j^+_C + j^-_C \right) \;,
\end{align}
\begin{align}
F^{\bar{C}}_{({\rm s})} &= -\left( \Delta\mu^{\bar{C}^+}_{+} + \Delta\mu^{\bar{C}^-}_{-} \right) \;, \nonumber \\
J_{\bar{C}}^{({\rm s})} &= \frac{1}{2} \left( j^+_{\bar{C}^+} - j^-_{\bar{C}^-} \right) \;, \\
F^{\bar{C}}_{({\rm a})} &= -\left( \Delta\mu^{\bar{C}^+}_{+} - \Delta\mu^{\bar{C}^-}_{-} \right) \;, \nonumber \\
J_{\bar{C}}^{({\rm a})} &= \frac{1}{2} \left( j^+_{\bar{C}^+} + j^-_{\bar{C}^-} \right) \;, \label{eq:C_bar_flow}
\end{align}
\begin{align}
F^{\rho}_{({\rm s})} & = \frac{1}{2}\left[ \vec{n}\cdot\left(\mathsf{T}^+_{\rm d}
                                 +\mathsf{T}^-_{\rm d}\right)\cdot\vec{n}- \Pi_+ - \Pi_- \right] \;, \nonumber \\
 J_{\rho}^{({\rm s})} &= \vec{n}\cdot \left( \vec{v}_+ - \vec{v}_- \right) \;, \\
F^{\rho}_{({\rm a})} & = \frac{1}{2}\left[ \vec{n}\cdot\left(\mathsf{T}^+_{\rm d}
                                 - \mathsf{T}^-_{\rm d}\right)\cdot\vec{n}- \Pi_+ + \Pi_- \right] \;,\nonumber \\
J_{\rho}^{({\rm a})} &= \vec{n}\cdot\left( \vec{v}_+ + \vec{v}_- -2\vec{v} \right) \; . \label{permeation}
\end{align}
The interpretations of the scalar currents and the pseudoscalar currents are rather obvious:
The scalar currents describe the heat absorbed, the material adsorbed, and the pseudoscalar
ones describe heat conduction, material permeation across the surface. 
\item[b)] The last two lines of Eq.~(\ref{s_production_perp_2}) describe the dissipation associated with
the hydrodynamic ``slip" between the surface and the two contacting bulk fluids.  The forces and currents
involved can be written as genuine vectors and pseudovectors,
\begin{align} 
F^{p}_{({\rm s}),\alpha} &= \vec{t}_\alpha\cdot\left(\vec{v}_+ + \vec{v}_--2\vec{v}\right) \;, \nonumber \\
J_{p}^{({\rm s}),\alpha} &= \frac{1}{2}\vec{n}\cdot\left(\mathsf{T}^+_{\rm d} - \mathsf{T}^-_{\rm d}\right)
                           \cdot\vec{t}^\alpha \;,\\
F^{p}_{({\rm a}),\alpha} &= \vec{t}_\alpha\cdot\left(\vec{v}_+ - \vec{v}_-\right) \;, \nonumber \\
J_{p}^{({\rm a}),\alpha} &= \frac{1}{2}\vec{n}\cdot\left(\mathsf{T}^+_{\rm d} + \mathsf{T}^-_{\rm d}\right)
                          \cdot\vec{t}^\alpha  \;.
\end{align}
\end{itemize}

\subsection{Constitutive relations}
The total entropy production associated with the dividing surface, the sum of
Eq.~(\ref{eq:s_production_tang}) and Eq.~(\ref{s_production_perp_2}), can now be expressed in terms
of the forces and currents defined above:
\begin{equation}
T\sigma_{s} \equiv T\sigma_{s,\parallel} + T\sigma_{s,\perp} = \sum_i J_i F^i \;,
\end{equation}
where $i$ runs over all the classes listed above.  The basis for developing constitutive relations
is our requirement that $T\sigma_{s}$ be positive definite, although it is developed for the
dividing surface in the Gibbs model system, which is not a physical surface.  This requirement ensures
the thermodynamic stability of any equilibrium state described by the theory; its validity will be
discussed in the last section of the paper.

The most general form of the constitutive relations is given by \cite{onsager,casimir45,kreuzer}
\begin{equation}
J_i=\sum_j\Omega_{ij}F^j\;,
\end{equation}
where $\Omega_{ij}$ are phenomenological parameters.  It immediately follows from Onsager's reciprocal
relations that $\Omega_{ij}$'s must satisfy a general property: $\Omega_{ij} = \Omega_{ji}$ if $J_i$
and $J_j$ have the same parity under time reversal, and $\Omega_{ij} = - \Omega_{ji}$ if $J_i$ and
$J_j$ have the opposite parity.  The positive definitiveness of $T\sigma_{s}$ is then ensured if
matrix $\Omega_{ij}$ is positive definite.  The anti-symmetric elements of $\Omega_{ij}$ require a few
more words.  It is not difficult to see that they do not contribute to the entropy production, in
other words, any possible couplings characterized by anti-symmetric $\Omega_{ij}$ are in fact reactive
couplings rather than dissipative couplings.  

In principle, any of the currents, $J_i$, may be driven by any of the forces, $F_j$, leading to the
so-called cross-coupling.  In other words, in the absence of pertinent symmetries or invariances,
all forms of cross coupling are possible.  However, if the physical description of the system is
invariant with respect to some or all of the orthogonal transformations, then the invariance will
eliminate certain cross-couplings.  For example, if the equilibrium shape of the membrane (or the
dividing surface) is symmetric with respect to rotations around any local normal vector, then scalar,
vector and tensor forces can only drive currents of the same type, leaving the coefficients
of all the related cross-coupling terms zero.  Similarly, if the physics of the system does not
distinguish one side of the membrane from the other, a pseudoscalar force can not drive a
scalar current and a scalar force can not a pseudoscalar current.  Consequently, the corresponding
$\Omega_{ij}$'s vanish.  If it turns out that these symmetries in practice are not exact, but almost
correct, then the corresponding cross-couplings will be weak.  An obvious approximation is to
discard those cross-coupling terms.  It may be worth noting, however, that symmetry properties
alone are not sufficient for identifying physically meaningful and relevant forces, currents,
and constitutive relations.  The general physics embodied in the entropy-production equations
and the specific physics of a system under consideration are necessarily needed.  

A few explicitly written constitutive relations may lead to some more intuitive understanding of
the general discussion given above.  For example, concerning the surface-intrinsic processes
described by Eq.~(\ref{heat_diffusion}), there are
\begin{align}
J_{s}^\alpha & =  K_{T} (-\partial^{\alpha}T) + \sum_{A\ne \w} \Omega_{s,A} (-\partial^{\alpha}\tilde{\mu}^{A}) 
\;, \nonumber \\
J_{A}^\alpha & =  \sum_{B\ne \w} \Omega_{A,B} (-\partial^{\alpha} \tilde{\mu}^{B}) + \Omega_{A,s}(-\partial^{\alpha}T)  
\;.
\end{align}
These two relations should be familiar to the reader, where $K_{T}$ may be interpreted as 
the effective heat conductivity of the surface, and $\Omega_{A,B}$'s are the effective dissipation
coefficients associated with diffusion.  Another example can be developed based on 
Eq.~(\ref{surface_viscosity}) and Eq.~(\ref{F_tensor}), and has the following form:
is the following:
\begin{align}
\vec{j}^\alpha_{p,{\rm d}} =&
\; -\eta_{\rm s}\left( g^{\alpha\gamma} g^{\beta\delta} + g^{\alpha\delta} g^{\beta\gamma}
                 -g^{\alpha\beta}g^{\gamma\delta} \right)
                \left(\partial_\gamma\vec{v}\cdot\vec{t}_\delta \right) \vec{t}_\beta \nonumber\\
& - \zeta_{\rm s}\left( \partial_\beta\vec{v} \cdot \vec{t}^\beta \right)\vec{t}^\alpha\;,\label{eq:surface_visc}
\end{align}
where the phenomenological constants $\zeta_{\rm s}$ and $\eta_{\rm s}$ may be interpreted as the only
relevant surface viscosity coefficients.  This constitutive relation is a two-dimensional analogue of
the familiar expression of viscous stress in a bulk fluid.  To be sure, the symmetry reasoning alone
would allow for the presence in the above constitutive relation of both the last antisymmetric part of
the force in Eq.~(\ref{F_tensor}) and a ``force" term of the form
$\left(\partial^\alpha\vec{v}\cdot\vec{n}\right)\vec{n}$.  The physical requirement that no dissipation
should be associated with uniform rotational motion of the whole system, however, renders those two
terms absent.  The latter term was allowed to be present in the work reported in Ref.~\cite{cai95}, though.  

The processes of transport of material across the surface are also of practical interest.
One of such processes is the so-called flip-flop process.  In the real world of membrane
systems it often occurs that this process is so slow as to be practically irrelevant for experimental
observations \cite{flipflop_1}.  However, situations where this process is relevant are also
encountered \cite{flipflop_2}, where the kinetic rate characterizing the process can be
determined experimentally.  In the framework of our theory, the simplest form of the
constitutive relation associated with the flip-flop of a chemical species $\bar{C}$ is
given by 
\begin{equation}
\xi^{K^*} = \Omega_{\updownarrow} \left[ - \left( \mu^{\bar{C}^+} - \mu^{\bar{C}^-} \right) \right]
\;,
\end{equation}
where the phenomenological parameter $\Omega_{\updownarrow}$ may be related to experimentally
determined flip-flop rate.  A related constitutive relation concerns the permeation of molecules
of a $\bar{C}$ type.  As it has been pointed out in connection with Eq.~(\ref{eq:C_bar_flow}),
the current of permeation, $J_{\bar{C}}^{({\rm a})}$, is a pseudoscalar, thus can be driven by
$F^{\bar{C}}_{({\rm a})}$ as well as by $F^{K^{*}} = - (\mu^{\bar{C}^+} - \mu^{\bar{C}^-} )$ defined
in Eq.~(\ref{eq:flipflop_2}). The corresponding constitutive relation may be written as
\begin{align}
J_{\bar{C}^{({\rm a})}} =& \; \Omega_{\bar{C},{\rm perm}}^{(1)} \left[ -\left( \Delta\mu^{\bar{C}^+}_{+} 
                                                           - \Delta\mu^{\bar{C}^-}_{-} \right) \right]\nonumber\\
&                        + \Omega_{\bar{C},{\rm perm}}^{(2)} \left[ -(\mu^{\bar{C}^+} - \mu^{\bar{C}^-}) \right]  
                        \;.                                                        
\end{align}
In contrast, in the constitutive relation characterizing the permeation of molecules of a
$C$-type, only the counterpart of the dissipative coefficient $\Omega_{\bar{C},{\rm perm}}^{(1)}$
exists.

The final example concerns the total permeation of material across the surface described by the
forces and currents defined in Eq.~(\ref{permeation}).  If it may be assumed that there is no cross
coupling between the scalar and the pseudoscalar quantities, then the following simple constitutive
relation may be written:
\begin{multline}
\label{darcy}
\vec{n}\cdot\left( \vec{v}_+ + \vec{v}_- -2\vec{v} \right)\\
= \Omega_{\rm perm} \frac{1}{2}\left[ \vec{n}\cdot\left(\mathsf{T}^+_{\rm d}
                                 - \mathsf{T}^-_{\rm d}\right)\cdot\vec{n}- \Pi_+ + \Pi_- \right] \;.
\end{multline}
The constant $\Omega_{\rm perm}$ is related to membrane permeability.  This relation would be
the familiar Darcy's law \cite{darcy}, but for the presence of the viscous stresses from the bulk
fluids.   

\subsection{From constitutive relations to equations of motion}
In the above, we have presented a framework of principles for developing constitutive
relations, the specific forms of which must depend on the particular physical system under
consideration.  We end this whole section with some comments on how the general theory may
be considered from an operational point of view.  

The complete set of equations of motion may be divided into two subsets: those that describe
the hydrodynamics of the bulk ``filler" fluids, i.e., Eq.~(\ref{bulk_consv}), and those
that describe the dynamics of the surface excess fields, i.e., Eq.~(\ref{surface_consv}).
Or, more specifically, Eq.~(\ref{eq:econs}) to Eq.~(\ref{eq:continuity}), together with the one
that governs the conformational dynamics, Eq.~(\ref{v_normal}), or Eq.~(\ref{surface_v}).
These two subsets of equations can only be solved if specific constitutive relations are
given or developed.  The constitutive relations associated with transverse transport 
processes provide the necessary coupling between the boundary values of the bulk hydrodynamic
fields and the surface dynamic fields, as Eq.~(\ref{darcy}) illustrates, for example.  They may
be viewed and used as boundary conditions for the bulk equations of motion.  An operational
strategy may then be to solve first the bulk equations of motion and express all the boundary
quantities of the bulk fluids appearing in the surface equations of motion in terms of the
surface dynamic fields, and then obtain an effective, closed set of equations of motion for
the surface dynamic fields.  As the focus of a study of dynamics of a membrane is naturally
on the surface fields, the effective equations should be the basis for analyzing the dynamics
of the membrane.

The number of the final, effective equations of motion, or the number of relevant dynamic fields,
associated with the dividing surface can be worked out in a general fashion.  Similar to the
case of a bulk fluid, relevant conserved variables are given by the surface densities of
excess energy, excess momentum, and numbers of excess particles, represented by $e$, $\vec{p}$
and $\{ n_{A} \}$, respectively, as in Eq.~(\ref{eq:econs}) to Eq.~(\ref{eq:continuity}).  An 
extra, non-conserved, and scalar, dynamic variable, however, becomes necessary for describing
the conformational aspect of membrane dynamics, as mentioned in the preceding paragraph
\footnote{In principle, the shape field, $\vec{R}$ is a vector, implying three scalar variables.
But, only one variable, corresponding to the motion of the membrane in its normal direction,
is the physically relevant one, as a consequence of the reparametrization invariance 
\cite{cai95}.}.  For
a reader with a particular interest in the linear analysis of equations of dynamics, the total
number of the relevant dynamic fields in principle determines the number of independent dynamic
modes that are implied by the effective equations of motion.   In practice, the number and the
specific nature of those modes that are relevant to, or observable in, experimental observations
of the dynamics of a particular system will have to be determined on a case-dependent basis.

Given its importance and accessibility for experimental observations \cite{french} the conformational
aspect of membrane dynamics on its own merits a comment.  Eq.~(\ref{v_normal}) may be considered
the corresponding equation of motion, and clearly, it is coupled to the other equations of motion.
That coupling is expected to be complicated in a general case.  In some limit cases, however,
an explicit form of the equation is available.  One of such limit cases, the case of membranes
made of single-component lipid-bilayers, has been explored earlier \cite{miao02} and will also be discussed
later in Sec.~\ref{subsec:1component} of this paper.  An effective equation for the conformational
dynamics in that limit case can be found in \cite{miao02}.

\section{The general theory in a few limit cases}
\label{sec:limitcases}
The formulation of the theory as presented in the previous sections is kept very general.
This feature is useful, seen from the formalistic point of view.  A few limit cases are,
however, interesting from the practical point of view.  In fact, another prediction of our theory,
which is in principle non-trivial, will be brought out more clearly in one of the limit cases.  In this
section we will discuss these limit cases.

\subsection{Multi-component membrane with no-slip boundary conditions}
\label{subsec:noslip}
In treatment of the bulk hydrodynamics of a viscous fluid in contact with a boundary, the boundary
condition that is usually imposed is the so-called no-slip boundary condition, where the bulk fluid
at the boundary is constrained to have the velocity of the boundary material.  

If this limit may be assumed as a valid one for the membrane-fluid system under our consideration, 
its implementation does not appear straightforward, due to the peculiar structural characteristics
of the membrane.  On the one hand, one type of molecular building blocks of the
membrane are amphiphilic molecules, which are distributed in two opposing monolayers, each in
contact with a different bulk fluid.  There is no physical reason to assume that the lipid molecules
forming the two different monolayers should have the same collective velocities, and in turn, no
reason to assume that the boundary velocities of the two bulk fluids are the same.  On the other hand,
another type of the molecular building blocks are transmembrane proteins, which span the whole thickness
of the membrane and which contribute to the dynamic interaction between the membrane and both of the
bulk fluids.  The question is, in other words, what surface velocities should be matched tangentially
with the boundary velocities of the two contacting bulk fluids.  The previously defined surface velocity
$\vec{v}$, which is the center-of-mass velocity of the whole surface, obviously is not the appropriate one.

We propose, as an answer to the question, that the no-slip boundary conditions be implemented by matching
the boundary velocities of the two bulk fluids with two in principle different linear combinations of the
velocities of all the different species associated with the dividing surface.  By use of the fact that
the collective velocity of a particular species $A$ is given by
$n_A\vec{v}_{A} \equiv \vec{j}_{A} = \left(j^\alpha_{A,{\rm d}}\vec{t}_\alpha + n_A\vec{v}\right)$,
the no-slip boundary conditions can be formulated as 
\begin{equation}
\label{hydromatching}
\vec{v}_\pm = \sum_{A} L^A_{\pm} \left(j^\alpha_{A,{\rm d}}\vec{t}_\alpha+n_A\vec{v}\right)\;.
\end{equation}
The $L^A_{\pm}$'s in the expression are phenomenological quantities, and may be interpreted as the
effective areas of the different species ``seen" by the bulk fluids.  Note that $\sum_A L^A_\pm n_A =1$
must be satisfied in order that the expression also hold for the case of uniform motion of all species.

Using this constraint together with Eq.~(\ref{eq:ddep}) leads to an alternative expression of the
no-slip boundary conditions
\begin{equation}
\label{no-slip}
\vec{v}_\pm = \vec{v} + \sum_{A\ne \w} {\tilde L}^A_{\pm} j^\alpha_{A,{\rm d}}\vec{t}_\alpha\;,
\end{equation}
where
\begin{equation}
\label{eq:L_tilde}
{\tilde L}^A_{\pm} \equiv L^A_\pm-\frac{m^A}{m^{\w}}L^{\w}_\pm\;.
\end{equation}
Eq.~(\ref{no-slip}) implies that $\vec{t}^\alpha\cdot \left( \vec{v}_+ - \vec{v} \right) $
and $\vec{t}^\alpha\cdot \left( \vec{v}_{-} - \vec{v} \right) $ are no longer independent of the
currents $\{ j^\alpha_{A,{\rm d}}, A \ne \w \}$, and it can then be used to eliminate 
$\vec{t}^\alpha\cdot \left( \vec{v}_{\pm} - \vec{v} \right)$ from those terms in
Eq.~(\ref{s_production_perp_2}) that contain them.  Collecting all the terms in the total entropy
production which contain $\{ j^\alpha_{A,{\rm d}}, A \ne \w \}$, including those in 
Eq.~(\ref{eq:s_production_tang}), finally leads to the identification of 
\begin{equation}
\label{force_diffusion}
F^A_\alpha = -\partial_\alpha \tilde{\mu}^A + \left( {\tilde L}^A_{+}\vec{n}\cdot\mathsf{T}^+_{\rm d}\cdot\vec{t}_\alpha
                                            -{\tilde L}^A_{-}\vec{n}\cdot\mathsf{T}^-_{\rm d}\cdot\vec{t}_\alpha\right)
\end{equation}
as the thermodynamic force conjugate to the diffusion current $J^\alpha_A$.

The terms contained in parenthesis in the above equation represent contributions from the 
two bulk fluids, and they underscore one of the specific predictions of our theory:  the exchange
of momentum between the surface and the bulk fluids is coupled to the dissipative processes in the
surface.  At the level of principles, this is a non-trivial prediction, and is made, to our 
knowledge, for the first time here.  In Section~\ref{sec:discussion}, an order-of-magnitude 
estimate of this new effect will be made with a view towards its experimental verification. 

\subsection{Single-component lipid bilayers}
\label{subsec:1component}
The case where the membrane is a bilayer composed of a single species of lipid molecules is also
worth discussing in connection with the earlier works dealing with such a system \cite{seifert93,miao02}.
In those earlier works, the description of the hydrodynamics of the membrane was established in
an intuition-based manner rather than from a systematic derivation.  In this section, we will
demonstrate that the earlier description can be formulated as a limit case of the formalism
developed here, thus putting the earlier description on a firmer basis.

In the earlier description, a number of limiting assumptions were made.  First, the temperature
was assumed to be uniform in the whole system; secondly, the membrane was assumed to be impermeable
to solution (water) molecules; thirdly, it was assumed that no transport of the lipid molecules
took place either between the bulk fluids and the membrane or between the two distinct monolayers
of the membrane; fourthly, it was assumed that the intramonolayer shear dissipations could be neglected,
but that the dissipation associated with the relative motion between the two monolayers was relevant;
finally, the Stokes condition was assumed, i.e. that the inertia of the membrane surface could also be
neglected.  Consistent with these assumptions, the single-component lipid bilayer was considered as a
composite of two systems ($+/-$), one for each monolayer.  Consequently, the description of the dynamics
became much simplified, consisting of two continuity equations 
\begin{equation}
\label{monolayer_mass_eq}
D_{t}n_{l_{\pm}} + D_{\alpha}j^{\alpha}_{l_{\pm}} \equiv  
D_{t}n_{l_{\pm}} + D_{\alpha} \left( n_{l_{\pm}}W^{\alpha}_{\pm} \right) = 0 \;,
\end{equation}
where $\vec{v}_{l_{\pm}} = W^{\alpha}_{\pm}\vec{t}_{\alpha} + \partial_{t}\vec{R}$ were the velocities
of the lipid molecules in the two monolayers, and by two additional equations which correspond to the
conservations of the monolayer momenta under the Stokes condition,
\begin{equation}
\label{eq:monoeom}
\vec{f}_{\rm rs}^\pm \pm \mathsf{T}^\pm\cdot\vec{n} + \vec{f}_{\rm m}^\pm = 0 \;. 
\end{equation} 
$\vec{f}_{\rm rs}^\pm$ represented the restoring force acting on each ($+/-$) monolayer, which should
be derived from the thermodynamic free energy, and $\vec{f}_{\rm m}^\pm$ was the force acting on one
monolayer by the other.  As a model for the intermonolayer dissipation, the following phenomenological
expression was employed
\begin{equation}
\vec{t}^{\alpha}\cdot \vec{f}_{\rm m}^\pm = -b\vec{t}^{\alpha} \cdot \left(\vec{v}_{l_{\pm}}-\vec{v}_{l_{\mp}}\right)
                                   = -b \left( W^{\alpha}_{\pm} - W^{\alpha}_{\mp} \right) \;,
\end{equation} 
where $b$ was the ``intermonolayer friction coefficient."  In order that the equations of the dynamics
be completely closed, intuition-based expressions were proposed for $\vec{f}_{\rm rs}^\pm$, which
satisfied 
\begin{equation}
\vec{f}_{\rm rs}^\pm\cdot\vec{t}_\alpha = -n_{l_{\pm}} \partial_\alpha\mu^{l_{\pm}} \;, 
\end{equation}
and 
\begin{equation} 
\vec{f}_{\rm rs}^+ + \vec{f}_{\rm rs}^- = \vec{f}_{\rm rs} \;.
\end{equation}

Eq.~(\ref{eq:monoeom}) can now be expressed in an alternative form, which will make a comparison with
the current formalism easier.  The sum of the two subequations yields the equation corresponding to
the momentum conservation for the whole bilayer, which is given by
\begin{equation}
\label{bilayer_stokes}
\vec{f}_{\rm rs} + \mathsf{T}^{+}\cdot\vec{n} - \mathsf{T}^{-}\cdot\vec{n} = 0 \; ;
\end{equation}
and the subtraction of the two subequations leads to the following expression 
\begin{align}
\label{eq:lingsubt}
& \frac{1}{2}\left(j^\alpha_{l_{+},{\rm d}}-j^\alpha_{l_{-},{\rm d}}\right)  \nonumber \\
& = \left(\frac{n_{l_{+}} n_{l_{-}}}{n_{l_{+}} + n_{l_{-}}} \right)^2\frac{1}{b} \;
  \bigg[\vec{n}\cdot\left(\frac{1}{n_{l_{+}}}\mathsf{T}^+
  - \frac{1}{n_{l_{-}}}\mathsf{T}^-\right)\cdot\vec{t}^\alpha\nonumber\\
&\phantom{ = \left(\frac{n_{l_{+}} n_{l_{-}}}{n_{l_{+}} + n_{l_{-}}} \right)^2\frac{1}{b} \;
  \bigg[}  -\partial^\alpha(\mu^{l_{+}} -\partial^\alpha\mu^{l_{-}} )\bigg]\;, 
\end{align}
where 
\begin{equation}
\label{diffusion_current}
j^\alpha_{l_{\pm},{\rm d}} \equiv j^\alpha_{l_{\pm}} - n_{l_{\pm}} \frac{j^\alpha_\rho}{\rho}
                          = j^\alpha_{l_{\pm}} - n_{l_{\pm}} 
                               \frac{j^\alpha_{l_+} + j^\alpha_{l_-}}{n_{l_{+}} + n_{l_{-}}} \;,
\end{equation}
obviously satisfying $j^\alpha_{l_{+},{\rm d}}=-j^\alpha_{l_{-},{\rm d}}$. 

When the same system is considered within the current formalism under the same limiting conditions,
the equations of dynamics for the system have the same formal expressions as Eq.~(\ref{monolayer_mass_eq})
and Eq.~(\ref{bilayer_stokes}), where $j^\alpha_{l_{\pm},{\rm d}}$ defined by Eq.~(\ref{diffusion_current})
represent the ``diffusion currents."  If, in addition to all the limiting conditions already mentioned,
the limiting, no-slip boundary conditions Eq.~(\ref{no-slip}) are also used, which in this simple case
reduce to 
\begin{equation}
\vec{v}_{+} = \vec{v}_{l_{+}} \;,  \quad \quad \vec{v}_{-} = \vec{v}_{l_{-}} \;,
\end{equation}
or equivalently to
\begin{equation}
{\tilde L}^{l_{+}}_{+} = L^{l_{+}}_{+} = \frac{1}{n^{l_{+}}} \;,  \quad \quad
{\tilde L}^{l_{+}}_{-} = - L^{l_{-}}_{-} = -\frac{1}{n^{l_{-}}} \;,
\end{equation}
the constitutive relation concerning the ``diffusion currents" is given by, following Eq.~(\ref{force_diffusion}),
\begin{align}
j^\alpha_{l_{+},{\rm d}} = - j^\alpha_{l_{-},{\rm d}}
& = \frac{1}{2} \left( j^\alpha_{l_{+},{\rm d}} - j^\alpha_{l_{-},{\rm d}} \right) \nonumber \\
& = \Omega_{\rm D} \bigg[\vec{n}\cdot\left( \frac{1}{n_{l_{+}}}\mathsf{T}^+_{\rm d}
                                  -\frac{1}{n_{l_{-}}}\mathsf{T}^-_{\rm d} \right)\cdot\vec{t}^\alpha\nonumber\\
&\phantom{ = \Omega_{\rm D} \bigg[} -\partial_{\alpha}(\mu^{l_{+}} - \mu^{l_{-}})  \bigg] \;.
\end{align}
It is obvious that, if the following identification is made,
\begin{equation}
\Omega_{\rm D}=\left(\frac{n_{l_{+}} n_{l_{-}}}{n_{l_{+}} + n_{l_{-}}}\right)^2\frac{1}{b} \;,
\end{equation}
the above constitutive relation becomes equivalent to \linebreak Eq.~(\ref{eq:lingsubt}).  The above equation 
makes it clear that the term ``diffusion current" within our formalism has a broader meaning
than its canonical one, in that it describes relative motion between species and does not have
to be associated with processes of molecular mixing.  

\subsection{Incompressible membrane and bulk fluids}
\label{subsec:incompressible}
Theoretical work dealing with a generic membrane system often in practice employs the assumption that
both the bulk fluids surrounding the membrane and the membrane itself are incompressible.  This 
assumption is justified in most cases.  With a view of applications of our general theory to those
practical situations, we in this subsection discuss the limit case of volumewise incompressible fluids
and membrane.

In what follows, we will model the limiting condition of incompressibility by assuming that the average
molecular volume associated with each molecular species in the membrane-fluid system remains constant
throughout the system, in other words, being insensitive to the spatial inhomogeneities in other
thermodynamic variables.  In addition we will only consider the case where another limiting condition
holds, that there is no transport of any molecules across the membrane.  This latter condition implies,
according to the definition of ``species" established earlier, that in the Gibbs-model description of
the system, two labelled species, $\bar{C}^+$ and $\bar{C}^-$, are associated with a single chemical
species $\bar{C}$.  It is not difficult to see that these conditions amount to two separate constraints
on the surface fields $\{ n_{\bar{C}^{\pm}} \}$.  For our purpose, the two constraints are expressed in
the following form
\begin{equation}
\label{eq:constraint}
n_{{0}^\pm} = B_\pm\left( \{n_{\bar{C}^{+} \ne 0^{+}},n_{\bar{C}^{-} \ne 0^{-}} \} \right)\;,
\end{equation}  
where $\bar{C} = 0$ represents the molecular species (water) forming the solvent.  

The immediate consequence of the constraints is that, among all the transverse currents of material 
transport,
$\displaystyle \left( \{ j^{\pm}_{\bar{C}^{\pm} \ne 0^{\pm}} \}, \; \rho^{\pm}\vec{n}\cdot (\vec{v}_{\pm}-\vec{v})
\right)$, 
two -- one for the ``$+$"-region and one for the ``$-$"-region -- must become dependent on the rest as
well as on the surface currents, $\{j^{\alpha}_{\bar{C}^{\pm},{\rm d}}\}$.  There are different ways
to remove these dependent currents from the entropy-production equation.  An obvious one is by direct
substitutions of their explicit dependences on the other currents.  Although straightforward, the
implementation is tedious and changes the formulistic structure of the derivations we have described
so far.  As a formulistically much more concise and simpler alternative, we have developed a different
approach to implement the constraints, which is sketched below.

In our approach, the following energy-density function is introduced first
\begin{align}
& e(n_{0^{+}},n_{0^{-}},\{ n_{\bar{C}^{+} \ne 0^{+}},n_{\bar{C}^{-} \ne 0^{-}} \} ) \nonumber \\
& = e_{\rm phys}(\{ n_{\bar{C}^{+} \ne 0^{+}},n_{\bar{C}^{-} \ne 0^{-}}\}) \nonumber\\
&\phantom{=}+ \lambda^{0^+}\left(n_{{0}^+}-B_+\right) 
                                                + \lambda^{0^-}\left(n_{{0}^-}-B_-\right)\;,
\end{align}
where $\displaystyle e_{\rm phys}(\{ n_{\bar{C}^{+} \ne 0^{+}},n_{\bar{C}^{-} \ne 0^{-}}\})$ represents
the physical energy-density function for the constrained system, and where $\lambda^{0^{\pm}}$ are two
multipliers to be determined.  It is obvious that, when the constraints are used,
$e(n_{0^{+}},n_{0^{-}},$ $\{ n_{\bar{C}^{+} \ne 0^{+}},n_{\bar{C}^{-} \ne 0^{-}} \} )$,
coincides with the physical function $e_{\rm phys}(\{ n_{\bar{C}^{+} \ne 0^{+}},n_{\bar{C}^{+} \ne 0^{+}}\})$
no matter what values the two multipliers take, and this fact will be taken advantage of in what follows.   

In the derivation of the entropy production the system is then treated {\em formally} as if it were not
constrained by letting $e(n_{0^{+}}, n_{0^{-}},\{ n_{\bar{C}^{+} \ne 0^{+}},n_{\bar{C}^{+} \ne 0^{+}} \} )$
play the role of the surrogate of the physical energy-density function.  The final expression of the
entropy production,
\begin{eqnarray}
\label{eq:s_production_total}
T\sigma_{s} & = &  -{j}^\alpha_{s,{\rm d}}\partial_\alpha T
                       - \vec{j}^\alpha_{p,{\rm d}}\cdot\partial_\alpha\vec{v}
                       - \sum_{A \ne \w} {j}^\alpha_{A,{\rm d}} \partial_\alpha \tilde{\mu}^A
                         \nonumber\\
         &  & - \sum_K \xi^K \Gamma_K+ \; {j}^-_s \left(T_- - T\right) - {j}^+_s \left(T_+ - T\right) \nonumber\\
&&             + \sum_{\bar{C} \ne 0} \left( {j}_{\bar{C}^{-}}^{-} \Delta\mu^{\bar{C}^{-}}_{-} 
                                       - {j}_{\bar{C}^{+}}^{+} \Delta \mu^{\bar{C}^{+}}_{+} \right) \nonumber \\
         &  & + \left(-\vec{n}\cdot\mathsf{T}^-_{\rm d}\cdot\vec{n} + \Pi_-\right) \vec{n}\cdot\left(\vec{v}_--\vec{v}\right)\nonumber\\
&&             +\left(\vec{n}\cdot\mathsf{T}^+_{\rm d}\cdot\vec{n} - \Pi_+\right)
             \vec{n}\cdot\left(\vec{v}_+-\vec{v}\right) \nonumber\\
         &  & - \left( \vec{n}\cdot\mathsf{T}^-_{\rm d}\cdot\vec{t}_\alpha \right)
                   \left[\, \vec{t}^\alpha\cdot \left( \vec{v}_--\vec{v} \right) \right]\nonumber\\
&&+ \left( \vec{n}\cdot\mathsf{T}^+_{\rm d}\cdot\vec{t}_\alpha \right)
                   \left[\, \vec{t}^\alpha\cdot \left( \vec{v}_+-\vec{v} \right) \right]  \;,         
\end{eqnarray}
thus looks formulistically identical to the sum of Eq.~(\ref{eq:s_production_tang})
and Eq.~(\ref{s_production_perp_2}), except that the multipliers $\lambda^{0^{\pm}}$ appear in the places
of the surface chemical potentials $\mu^{0^{\pm}}$ in
\begin{align}
 \Delta\mu^{\bar{C}^{\pm}}_{\pm} \equiv& \left(\mu^{\bar{C}^{\pm}}_{\pm}-\frac{m^{\bar{C}}}{m^{0}}\mu^{0^\pm}_{\pm}\right)\nonumber\\
&                                  -\left( \mu^{\bar{C}^{\pm}} 
                                        -\frac{m^{\bar{C}}}{m^{0}} \lambda^{0^\pm} \right) \;, \quad  \bar{C} \ne 0  \;,  \\
 \Pi_\pm  \equiv &\; \frac{\rho^\pm}{m^{0}} \bigg[ \frac{1}{2}m^{0} \left(\vec{v}_\pm-\vec{v}\right)^2
             + (\mu^{0^\pm}_\pm + \frac{1}{2}m^{0}\vec{v}^2_\pm )\nonumber\\
&\phantom{\frac{\rho^\pm}{m^{0}} \bigg[}             - (\lambda^{0^\pm} + \frac{1}{2}m^{0}\vec{v}^2) \bigg] \;.
\end{align}

Eq.~(\ref{eq:s_production_total}) still contains all currents, both independent ones and the corresponding
dependent ones.  The freedom in choosing the values of the multipliers $\lambda^{0^{\pm}}$ when the
constraints are satisfied now means that, given a particular choice of the independent currents, the
corresponding dependent ones can then be eliminated from Eq.~(\ref{eq:s_production_total}) by choosing
values for $\lambda^{0^{\pm}}$ accordingly.  For example, for the choice where
$\displaystyle \left( \{ j^{\pm}_{\bar{C}^{\pm}},\bar{C}\ne 0\} \right) $ are used as the independent
transverse currents, the values of $\lambda^{0^{\pm}}$ must be chosen such that 
\begin{equation}
\label{eq:multiplier_choice}
\vec{n}\cdot\mathsf{T}^\pm_{\rm d}\cdot\vec{n} - \Pi_\pm=0 \;.
\end{equation}
This choice renders both the fourth and the fifth lines in Eq.~(\ref{eq:s_production_total}) zero, thus
eliminating the corresponding dependent currents $\rho^{\pm}\vec{n}\cdot (\vec{v}_{\pm}-\vec{v}) $.

The values of $\lambda^{0^{\pm}}$ chosen as such, together with the constitutive relations which
can be derived consequently, should then be used in conjunction with equations of motion where the
constraints are explicitly satisfied, in order that the whole formulation of the dynamics form
a self-consistent and closed one.

The implications of Eq.~(\ref{eq:multiplier_choice}) merit a short comment.  It follows from the
the definition of $\Pi_\pm$ that the values of $\lambda^{0^{\pm}}$ are linearly related to the
chemical potentials of the bulk solvent molecules, which in turn are related to the pressures
in the bulk fluids.  Thus, bulk-pressure gradients tangential to the surface,
$\vec{t}_{\alpha} \cdot \vec{\nabla}p_{\pm}$, are translated by Eq.~(\ref{eq:multiplier_choice})
into tangential gradients of $\lambda^{0^{\pm}}$ and can thereby drive diffusion currents within
the surface in principle, as implied by Eq.~(\ref{force_diffusion}).  The practical relevance
of this mechanism can, however, be argued to be insignificant.  A quick estimate reveals that the
tangential gradients of $\lambda^{0^{\pm}}$ are given roughly by
$ w^{0}\vec{t}_{\alpha} \cdot \vec{\nabla}p_{\pm}$, where $w^{0}$ denotes the molecular volume
of the solvent.  When compared with the contributions in Eq.~(\ref{force_diffusion}) from the
bulk shear stresses, this effect is suppressed likely by the ratio between a molecular length
and the wavelength characterizing the bulk-pressure gradients, thus probably insignificant.  
This suppression means that, although the excess quantity of the solvent molecules associated
with the surface may not be negligible, its effect on the motion of other surface-related
molecular species may be assumed to be insignificant.  Our understanding here justifies, therefore,
the use of this assumption in the previous works on membrane hydrodynamics \cite{miao02,onuki93}.   

A general remark may also be made to conclude this subsection.  Although our method of constraint
implementation has been applied in the specific limit case of systems consisting of incompressible
fluids and incompressible membranes, it may also be applied to cases where constraints are different.
The difference in the use of the method will only result from the different details of the constraint
functions used in Eq.~(\ref{eq:constraint}).

\section{Discussions}
\label{sec:discussion}
In the previous sections, a formulation of hydrodynamics of the membrane-fluid systems has been
presented in a very general form, both in terms of equations of motion based on the basic conservation
laws and in terms of derived constitutive relations.  Although an application of the general theory
to any specific system is beyond the scope of this paper, we will, nevertheless, discuss in 
this section a number of general issues which must be dealt with in applications and which
will also make the connection between the theory and experimental situations visible.  For the
following discussion we reiterate that the theory is only an effective one and that its purpose
is to give a mesoscopic/macroscopic-scale description of the membrane dynamics, which can be
connected to experimental descriptions on the corresponding scales.

The first issue concerns the structure and the content of the theory.  From the way the theory
is constructed, it is clear that its description of the motion in the regions close
to the membrane interface is, by design, not correct qualitatively.  The obvious questions are, then,
how the quality of the theory can be controlled, and what criteria should be used as measures
of the quality of the theory.  The means of the control is already contained in the theory itself
and rests on tuning the set of kinetic coefficients associated with the various constitutive
relations.  Whenever the theory should be applied, it is assumed that the hydrodynamic conditions
at the boundaries of the entire system are controlled by macroscopic means and are, therefore, known.
Given those boundary conditions, and provided that the dynamical behaviour of the system in the
regions sufficiently far away from the membrane (or the dividing surface), or the ``bulk behaviour,"
is known already, or may be known, by ``measurements," the set of kinetic coefficients can then be
tuned in the process of solving the equations of motion under the given boundary conditions, such
that the quantitative description of the ``bulk behaviour" given by the theory matches that given
by the measurements to some desired degree of agreement.  Moreover, in the theory, a few quantities
pertaining to the interfacial region can be calculated, 
\begin{enumerate}
\item $\int_{\bar{\Sigma}} dV\;{\bar x}(\vec{r},t)\;$,
\item $\int_{\bar{B}_\pm^\alpha}d{\vec{\bar{A}}}\cdot
     \left[ \vec{\bar{J}}_X - \bar{x}\partial_t\left(\vec{R}+h\vec{n}\right) \right] \;$,
\end{enumerate}
where $\bar{\Sigma}$ is the familiar volume element already used in Section~\ref{subsec:basicequation}
and
\begin{multline}
{\bar{B}_\pm^\alpha} = \Big\{\vec{R}(\bar{\uc}^1,\bar{\uc}^2,t)
                         +h \vec{n}(\bar{\uc}^1,\bar{\uc}^2,t)\;\;\Big|\\
\uc^{3-\alpha}-\Delta\uc^{3-\alpha}/2\leq \bar{\uc}^{3-\alpha}\leq \uc^{3-\alpha}
                         +\Delta\uc^{3-\alpha}/2\;,\\
-\epsilon^-\leq h \leq \epsilon^+\;,\;\;\bar{\uc}^\alpha=\uc^\alpha\pm\Delta\uc^\alpha/2
\Big\}\;
\end{multline}
represent the side surfaces of $\bar{\Sigma}$ and $d{\vec{\bar{A}}}$ is the area element on this surface
times its outward-pointing normal vector.   The counterparts of these quantities, or at least some of
them, in the real physical system may also be obtained from measurements.  One example would be 
the measurements of distribution and motion of membrane proteins within the membrane by the use
of modern techniques such as Fluorescence Confocal Microscopy \cite{fluorescence_1} and Fluorescence
Correlation Spectroscopy \cite{fluorescence_2}.  Such experimental information on the ``surface behaviour"
of the system, whenever available, must also be used to tune the theoretical kinetic coefficients,
thereby to control the quality of the theory.

Besides tuning the kinetic coefficients such that the quantities labelled 1. and 2. above are identical
for the Gibbs model and the experiment, the membrane parameters should also be tuned such that the
position of the shape $\vec{R}$ matches the position of the dividing surface in the experiment.

The second issue is a general one concerning the tuning of the kinetic coefficients.  It has been
a requirement, made in Section 6.2, that the entropy production in the theory be positive
definite in order that the equilibrium states of the theory be stable.  One consequence of this
requirement, among others, is that kinetic coefficient $\Omega_{ii}$ for each $i$ is positive. 
It turns out that in certain situations, this stability requirement can only be fulfilled by
an approximation in the tuning of the kinetic coefficients.  The following example of permeation
of molecules across a membrane illustrates the point.     

Consider the following simplified picture of a real (in contrast to the Gibbs model) membrane-fluid
system. The membrane is taken to be a homogeneous slab of material of thickness $2d$.  A certain
molecular species has different diffusion constants, $D'$ and $D$, in the membrane region and in
the two bulk fluids, respectively.   At a distance $L_1$ from the membrane slab, there
is a reservoir of the molecules with a constant chemical potential $\mu_1$, and similarly on
the other side, a reservoir of chemical potential $\mu_2 < \mu_1$ is placed at a distance $L_2$
from the membrane slab. 

The quantity of interest is the steady-state molecular flux, $j^{\rm real}$, between the two
reservoirs. It is easy to work out that 
\begin{equation}
j^{\rm real} = \frac{D D'}{D'(L_1+L_2)+2d D}\left(\mu_1-\mu_2\right)\;.
\end{equation} 

In the theory, the real system is replaced by a Gibbs model system: two bulk fluids, where the
molecular species has diffusion constant $D$, separated by an infinitely thin dividing interface
with an effective kinetic coefficient $\Omega$ for permeation of the molecules, which must be tuned.
The two reservoirs of chemical potentials $\mu_1$ and $\mu_2$ are then at distances $L_1+d$ and $L_2+d$
from the interface.  In order that the steady-state molecular flux be reproduced in the model system,
it is necessary that the boundary chemical potentials of the bulk fluids at the interface be different,
denoted by $\mu^{+}$ and $\mu^{-}$.  The steady-state flux, $j^{\rm model}$, can then be expressed as
\begin{align}
j^{\rm model} &= -D\frac{\mu_2-\mu^+}{L_2+d} = -\Omega\left(\mu^+-\mu^-\right)\nonumber\\
&=-D\frac{\mu^--\mu_1}{L_1+d} \;.
\end{align}
From this equation an alternative expression of $j^{\rm model}$ can be obtained 
\begin{equation}
\label{eq:model_flux}
j^{\rm model} = \frac{D\Omega}{\Omega(L_1+L_2+2d)+D}\left(\mu_1-\mu_2\right) \;.
\end{equation}

In order that the requirement, $j^{\rm model} = j^{\rm real}$, be fulfilled, 
$\Omega$ must be set to be
\begin{equation}
\label{eq:Omega}
\Omega = \frac{D D'}{2d(D-D')} \;.
\end{equation}

Eq.~(\ref{eq:Omega}) immediately illustrates a problem of the theory in the case where the molecules
diffuse faster than they do in the bulk fluids, i.e. $D'>D$: on the one hand, the effective
kinetic coefficient $\Omega$ must be given a negative value; on the other hand, any equilibrium
state becomes unstable in the theory when $\Omega$ is negative.

This problem does not, however, render the theory inapplicable to the case. A reasonable approximation
can still be made, which involves setting $\Omega$ to infinity. The chemical boundary condition for
the bulk fluids at the dividing surface thus becomes $\mu^{+} = \mu^{-}$.  The error introduced in
this approximation has an upper bound, which can be easily calculated:
\begin{align}
\frac{j^{\rm real}-j^{\rm model}}{j^{\rm real}} &=
\frac{2d}{L_1+L_2+2d}\cdot\frac{D'-D}{D'}\nonumber\\
& < \frac{2d}{L_1+L_2} \;.
\end{align}
Clearly, the theory should still give quantitatively reasonable result if $d$ is much smaller than
$L_{1}$ or $L_2$, or in a more general situation, if the length scale associated with the chemical
gradient in the system is much larger than the microscopic thickness of the membrane
\footnote{The negative permeability is a direct consequence of invoking the Gibbs model in cases
where $D'>D$.  An alternative formulation may also be envisioned, as suggested by one of the
referees of this manuscript.  In this formulation, the thickness of the membrane would be taken
to zero without corresponding ``compensation" in the theory for the effects that are associated
with the finite thickness of the membrane.  As a consequence, the kinetic coefficient associated
with transmembrane permeation stays positive in all cases.  This zero-thickness treatment is
obviously an approximation, as is our suggested solution to the problem of negative permeability.
The errors introduced due to this approximation may be comparable to those associated with our
formulation.}.

The last issue to be touched upon here is the question of how the various kinetic coefficients in
the effective theory may be related to experimental data on the corresponding transport processes.  
The precise answer to the question depends on the nature of the experiments that are done.  For
example, if a permeation experiment is done in the way described above, where the molecular flux
is measured, then an effective kinetic coefficient $\Omega$ can be derived based on Eq.~(\ref{eq:model_flux}).
Unfortunately, very few experiments have been done so far on the investigations of membrane-involved
transport processes; concrete situations where this issue can be addressed specifically are hard to 
find.  We hope, however, that our theory may provide some guidelines for designing useful experiments
regarding membrane-involved transport processes and for interpreting data obtained. 

It would add strength to a general theory such as the one presented in this paper if it makes
conceptually non-trivial and experimentally relevant predictions of new physical effects.  Indeed,
our theory does make predictions that are conceptually non-trivial, among which the following 
constitutive relation -- the simplest one that can be written down based on Eq.~(\ref{force_diffusion}),
\begin{multline}
\label{eq:new_prediction}
j_{A,{\rm d}}^{\alpha} = \Omega_{A,A} g^{\alpha \beta}F^A_\beta 
= \Omega_{A,A}g^{\alpha \beta} \Big[ -\partial_\beta \tilde{\mu}^A \\
                              + \left( {\tilde L}^A_{+}\;\vec{n}\cdot\mathsf{T}^+_{\rm d}\cdot\vec{t}_\beta
                              -{\tilde L}^A_{-}\;\vec{n}\cdot\mathsf{T}^-_{\rm d}\cdot\vec{t}_\beta \right)
                              \Big] \;,
\end{multline}
is the most notable.  An order-of-magnitude estimate of the effect predicted by Eq.~(\ref{eq:new_prediction}),
namely, that the viscous stresses exerted by the bulk fluids on the membrane can drive the surface 
diffusion processes in the membrane, may shed some light on the experimental relevance of this
theoretical prediction.    

Consider the case where a membrane is composed of a lipid species and a protein species embedded in the
lipid bilayer.  The subject of our interest is the motion of the embedded protein.   Assume that the
physical conditions are such that there exists a surface gradient of the protein concentration in the
membrane, sufficient to drive appreciable protein diffusion.  In addition, the membrane is subjected
to a shear flow induced in one or both of the bulk fluids.  For an order-of-magnitude estimate it suffices
that we assume near-plane geometry for the membrane and that we include in the chemical potential of the
protein only the contribution from entropy of mixing, i.e.
\begin{equation}
\mu^{\rm p}=kT\ln n_{\rm p}\;.  
\end{equation}
It follows immediately then that
\begin{equation}
\partial_{\alpha}\mu^{\rm p} \sim \frac{kT}{n_{\rm p}}\frac{\Delta n_{\rm p}}{\Delta l} \;,
\end{equation}
where $\Delta l$ denotes the wavelength characterizing the concentration gradient.  
The contribution from the bulk shear stress can also be estimated, based on Eq.~(\ref{eq:new_prediction}),
to be
\begin{equation}
{\tilde L}^{\rm p} \cdot \; \left( \eta\frac{\Delta v}{\Delta l} \right) \;,
\end{equation}
where ${\tilde L}^{\rm p}$ is in order of magnitude similar to ${\tilde L}^{\rm p}_{\pm}$ defined
by Eq.~(\ref{eq:L_tilde}) with the species index $O$ referring to the lipid, but in explicit expression
different from ${\tilde L}^{\rm p}_{\pm}$ \cite{michael_thesis}.

For the newly predicted effect to be comparable to the effect associated with the chemical 
gradient,  the variation of the velocity characterizing the bulk shear flow should be
\begin{equation}
\Delta v \sim \frac{kT}{\eta {\tilde L}^{\rm p}}\frac{\Delta n_{\rm p}}{n_{\rm p}}\;.
\end{equation}
Taking $\eta_0=10^{-3}\,{\rm Pa}\cdot{\rm s}$, the shear viscosity of water at room temperature
($kT=4\cdot 10^{-21}{\rm J}$) and setting
${\tilde L}^{\rm p} \sim 500${\AA}$^{2} = 5\cdot 10^{-18}\,{\rm m^2}$, a reasonable value for
the cross-sectional area of a typical transmembrane-protein mole-\linebreak cule, we get
\begin{equation}
\label{eq:shearest}
\Delta v \sim \left( \frac{\Delta n_{\rm p}}{n_{\rm p}}\cdot \frac{\eta_0}{\eta} \right) \frac{{\rm m}}{{\rm s}}\;.
\end{equation}
Based on the above expression, it may be concluded that it is very plausible that protein diffusion
induced by bulk shear flows can be observed experimentally, if the shear viscosity (or viscosities)
of the bulk fluids can be increased.   

Finally, we conclude this paper with our outlook, from the stand point of this work, on study of
non-equilibrium dynamics of membrane systems.  It is our belief that non-equilibrium behaviour of
membrane systems must be investigated, described and understood.  It is then our hope that the
theory presented in this paper will contribute to that study as a general framework, from which
applications to specific concrete problems can grow.   Compared to more intuition-based approaches
to formulating descriptions of non-equilibrium dynamics of membrane systems, the present formal
approach has the virtue that it avoids guesswork and ensures that relevant effects or mechanisms
will be included and correctly described in the equations of motion and in the boundary conditions. 
To be sure, not all of the constitutive relations rendered possible in the general theory are
relevant in a specific application to a specific system.  But, a systematic analysis of what
is relevant or what is not can be made within this framework.  If assumptions are made, then
the conditions under which the assumptions are valid can also be made clear.  

Indeed, we have shown in Sec.~\ref{subsec:1component} how the general theory applies to the simple
case of non-equilibrium dynamics of single-component membranes and under what limiting conditions
the general theory reduces to the earlier descriptions \cite{seifert93,miao02}.  Another obvious
case to apply the general theory to is the system of active membranes studied already both
experimentally and theoretically \cite{french}.  That application will be presented in its full
detail in a forthcoming paper \cite{active}.   In the context of this paper, however, a question
naturally arises concerning that application: does the general theory lead to any results that
are qualitatively different from the results derived from the theory developed in a more intuitive
way in Ref.~\cite{french}? The answer, briefly stated, is yes.  The general theory differs conceptually
from the earlier theory in terms of several new elements: first, the inclusions of the bulk viscous
stresses in the driving force for the transverse permeation processes, as stated in Eq.~(\ref{darcy}),
as well as in the driving forces for lateral diffusion processes, as described in
Eq.~(\ref{eq:new_prediction}); secondly, a new proposal for the conditions of matching between
the bulk and the surface hydrodynamic velocities, as formulated in Eq.~(\ref{hydromatching});
finally, the prediction of a mechanism where surface-tangential gradients of the bulk pressures
can drive lateral diffusions, as discussed in Sec.~\ref{subsec:incompressible}.  Under the physical
conditions that are assumed to hold for the active membranes considered, in particular, the condition
that the membranes may be considered impermeable, the first element may be ignored in practice.
It is much harder to ignore the other two.  In fact, the analysis in Ref.~\cite{active}
seems to suggest that the last element may contain a plausible theoretical interpretation
of certain experimental observations on diffusion dynamics of the active proteins in the
membranes.   

The general theory may prove to have another potential, which is also demonstrated in its 
application to the active membranes \cite{active}.  The potential derives from the underlying
philosophy of the theory: that physical effects associated with the microscopic ``bulk"
regions in contact with the membrane can be reformulated in terms of surface excess quantities,
if the microscopic details are not the focus of a description.  The potential is fully taken 
advantage of in the application and renders technically tractable a calculation that would
otherwise be prohibitively difficult, if not completely impossible.       

The current theory, despite being sufficiently general, is still limited in its scope of
coverage of systems or phenomena.  We would like to think that its philosophy will lend
itself to future development of even more general theories for describing more complex
non-equilibrium phenomena.  

\vspace{1cm}
\noindent
{\small
ML acknowledges the financial support from University of \linebreak Southern Denmark in the form of
a Ph.D. fellowship.  PLH and LM would like to thank the Danish National Research
Foundation for its financial support in the form of a long-term operating grant 
awarded to The MEMPHYS-Center for Biomembrane Physics.  The authors are grateful to 
Tove Nyberg for her technical assistance on the figures. }

\appendix
\section{Transforming density fields}
\label{se:trans}
This short appendix is included to describe a technical point which may sometimes be encountered 
in a specific calculation based on the theory.  In certain situations, it can be convenient to do
calculations by using different linear combinations of the physical density fields, $n_{A}$'s, rather
than the fields themselves.  An example is that one may want to use, as the ``working" density fields,
those combinations that have a certain parity when the chosen direction of the normal director to the
dividing surface is reversed.\footnote{See for example Ref.~\cite{miao02}, where the so-called ``sum field"
and the ``difference field" are used.}  

A general linear transformation of the $n_{A}$'s takes the following form,
\begin{equation}
{n'}_A=\sum_B n_B M^{B}_{\phantom{B}A}\;,
\end{equation}
where $M^{B}_{\phantom{B}A}$ is a constant matrix with $\det M^{B}_{\phantom{B}A}\ne 0$.  If ${n'}_A$'s
are chosen to be the ``working" density fields, equations governing the dynamics of the ${n'}_A$-fields
become the equations to be used in the calculation.  In order that the physics of the problem remain
invariant under the transformation, other related fields, constants and currents must also be 
transformed.  Specifically, the continuity equations governing the dynamics of ${n'}_A$'s will in fact
have the same mathematical forms as those for the $n_{A}$-fields, but, different effective masses,
${m'}^A$'s, must be associated with the ${n'}_A$-fields, which are given by the transformation
\begin{equation}
{m'}^A=\sum_B (M^{-1})^A_{\phantom{A}B}m^B\;,
\end{equation} 
where $(M^{-1})^A_{\phantom{A}B}$ is the inverse of $M^{B}_{\phantom{B}A}$, i.e.
\begin{equation}
\sum_B M^{A}_{\phantom{A}B}(M^{-1})^B_{\phantom{B}C}=\delta^A_{\phantom{B}C} \;.
\end{equation}
Moreover, the mass currents associated with the ${n'}_A$-fields must also be the transform
of the physical currents $j^\alpha_A$, $j^\alpha_{A,{\rm r}}$, $j^\alpha_{A,{\rm d}}$ and $\nu_{A,K}$
by the matrix $M^{B}_{\phantom{B}A}$, while the associated chemical potentials are the results
of the transformation of $\mu^A$ by the matrix $(M^{-1})^A_{\phantom{A}B}$.  

The transformations of quantities such as $\tilde{\mu}^{A}$ and kinetic constants $\Omega_{A,B}$,
where the species indices $A$ and $B$ do not take value $\w$, are in general more complicated.
If the following conditions,  
\begin{equation}
M^{\w}_{\phantom{1}A}=0\;,\quad M^{A}_{\phantom{1}\w}=0\;,
\end{equation}
are satisfied, however, $\Omega_{A,B}$ still transform according to a relatively simple rule
\begin{equation}
{\Omega'}_{AB}=\sum_{C,D}\Omega_{CD}M^{C}_{\phantom{C}A}M^{D}_{\phantom{D}B}\;.
\end{equation}

It may be worth mentioning that the convention which has been used so far of placing the species-label
index $A$ as either a superscript or a subscript has been chosen such that it conforms with the general
convention of index contraction in matrix algebra.

\end{document}